\documentclass{aastex6}
\usepackage[hyperref]{} 

\usepackage{epsfig}
\usepackage{psfig}
\usepackage{color}
\usepackage{graphicx}

\newcommand{\Bvec}{{\mathbf B}}
\newcommand{\bvec}{{\mathbf B}}

\newcommand{\be}{\begin{equation}}
\newcommand{\ee}{\end{equation}}
\newcommand{\bea}{\begin{eqnarray}}
\newcommand{\eea}{\end{eqnarray}}
\newcommand{\beax}{\begin{eqnarray*}}
\newcommand{\eeax}{\end{eqnarray*}}
\newcommand{\ea}{\end{array}}
\newcommand{\bed}{\begin{description}}
\newcommand{\ed}{\end{description}}
\newcommand{\blc}{\begin{list}{$\circ$}{}}
\newcommand{\blb}{\begin{list}{$\bullet$}{}}
\newcommand{\el}{\end{list}}
\newcommand{\een}{\end{enumerate}}

\def\gsim{\,\lower3pt\hbox{$\sim$}\llap{\raise2pt\hbox{$>$}}\,}
\def\lsim{\,\lower3pt\hbox{$\sim$}\llap{\raise2pt\hbox{$<$}}\,}

\newcommand{\hatr}{\,\hat{\mathbf r}}

\newcommand{\mxcm}{\ensuremath{~{\rm  Mx}\cdot {\rm cm}^{-2}}}

\renewcommand{\aap}{    {\it Astron. Astrophys.}}

\renewcommand{\apj}{    {\it Astrophys. J.}}
\renewcommand{\apjl}{   {\it Astrophys. J. Lett.}}

\renewcommand{\jgr}{    {\it J. Geophys. Res.}}

\renewcommand{\solphys}{{\it Solar Phys.}}

\begin{document}
\tracingmacros=2


\title{Active Region Emergence \& Remote Flares} 

\author{Y. Fu\altaffilmark{1}}
\affil{Department of Physics and Astronomy, Rutgers University, 136 Frelinghuysen Rd, Piscataway, NJ 08854}
\email{yxfu@physics.rutgers.edu}
\and
\author{Brian T.~Welsch\altaffilmark{2}}
\affil{Natural \& Applied Sciences, University of Wisconsin - Green
  Bay, 2420 Nicolet Dr., Green Bay, WI 54311}
\email{welschb@uwgb.edu}

\altaffiltext{1}{Department of Physics, University of California,
366 LeConte Hall, MC 7300, Berkeley, CA 94720-7300} 
\altaffiltext{2}{Space Sciences Laboratory, 7 Gauss Way, University of
California, Berkeley, CA 94720-7450}

%
%
%
%



\shortauthors{Fu \& Welsch}
\shorttitle{Active Region Emergence \& Remote Flares}


\begin{abstract}
We study the effect of new emerging solar active regions on the
large-scale magnetic environment of existing regions.  We first
present a theoretical approach to quantify the ``interaction energy''
between new and pre-existing regions as the difference between (i) the
summed magnetic energies of their individual potential fields and (ii) the
energy of their superposed potential fields.
%
%
We expect that this interaction energy can, depending upon the
relative arrangements of newly emerged and pre-existing magnetic flux,
indicate the existence of ``topological'' free magnetic energy in the
global coronal field that is independent of any ``internal'' free
magnetic energy due to coronal electric currents flowing within the
newly emerged and pre-existing flux systems.
We then examine the interaction energy in two well-studied cases
of flux emergence, but find that the predicted energetic perturbation
is relatively small compared to energies released in large solar
flares.  Next, we present an observational study on the influence of
the emergence of new active regions on flare statistics in pre-existing
active regions, using NOAA's Solar Region Summary and GOES flare
databases.  As part of an effort to precisely determine the emergence
time of active regions in a large event sample, we find that emergence
in about half of these regions exhibits a two-stage behavior, with an
initial gradual phase followed by a more rapid phase.
Regarding flaring, we find that the emergence of new regions is
associated with a significant increase in the occurrence rate of X-
and M-class flares in pre-existing regions. This effect tends to be
more significant when pre-existing and new emerging active regions are
closer.
Given the relative weakness of the interaction energy, this effect
suggests that perturbations in the large-scale magnetic field, such as
topology changes invoked in the ``breakout'' model of coronal mass
ejections, might play a significant role in the occurrence of some
flares.
\end{abstract}
\tracingmacros=0

\section{Introduction}
\label{sec:intro}

The emergence of new active regions is expected to significantly alter
the magnetic environment of pre-existing active regions (PEARs).  With
any modification of radial flux distribution on the photosphere, the
global potential magnetic field --- the unique, current-free magnetic
field matching the same photospheric radial field --- is
changed. Since the potential field is the hypothetical lowest-energy
state of the global magnetic field above the photosphere, the
emergence of a new region might create a new lowest-energy magnetic
configuration for PEAR fields, even if the actual magnetic field
within PEARs remains essentially unchanged.  The amount and
distribution of free magnetic energy (e.g., \citealt{Forbes2000,
  Welsch2006}) can therefore change, and an increase in free energy
might ``activate'' PEARs to produce flares or coronal mass ejections
(CMEs). 

This idea is not new. For instance, \citet{Heyvaerts1977} suggested
long ago that the emergence of a new flux system will generally induce
current sheets to flow on the separatrices between pre-existing flux
systems and the new one, which should lead to magnetic reconnection
and energy release. \citet{Pevtsov2004} suggested this reconnection
may play a role in coronal heating, and cite as evidence observations
of a ``diffuse cloud'' of enhanced soft X-ray emission (distinct from
loop-like emission) in a large halo (out to 500$''$) surrounding
a new region.  Also, observers have reported evidence of
emergence-induced disruption of pre-existing filaments (e.g.,
\citealt{Bruzek1952, Feynman1995, Wang1999,
  Balasubramaniam2011b}).
This process has also been modeled in case studies. For example,
\citet{Longcope2005a} and \citet{Tarr2012} constructed detailed models
of reconnection between a new region and a pre-existing one.  More
recently, \citet{Tarr2014} investigated the steady, ``quiescent''
reconnection that increased the magnetic flux linking a new and
pre-existing flux system.


The consequences of AR emergence on PEARs can include destabilizing
the coronal field on global scales.  For instance, a trigger of the
``breakout'' coronal mass ejection process \citep{Antiochos1999a} is
the removal of overlying, ``strapping'' fields that confine a
low-lying sheared field or flux rope.  If reconnection to a new active
region were to transfer flux that comprised strapping fields from the
old region to a new one, a CME might ensue.  
Observationally, \citet{Feynman1995} found that pre-existing quiescent
filaments near new active regions were more likely to erupt in the
days following the emergence than filaments not near new regions.
Further, they found eruptions to be still more likely when ``the new
flux was oriented favorably for reconnection with the preexisting
large-scale coronal arcades.''
More recently, \citet{Schrijver2011b} investigated an episode of
global-scale magnetic restructuring in a series of seemingly linked
flares and CMEs, in which one event triggered other, ``sympathetic''
events in distant active regions.
In another recent study, \citet{Schrijver2015} found $>$ 1-$\sigma$
increases (assuming Poisson statistics) in the rates of flares and
eruptions from distant regions ($> 20^\circ$) within four hours after
large flares --- of GOES class M5 or higher --- from May 2010 through
2014.
They interpret increased rates of distant events as evidence that
  remote flares/eruptions can be triggered by global-scale magnetic
  perturbations from large flares.
We expect that the emergence of new active regions can also
induce major changes into the large-scale structure of the corona, so
might play a role in global-scale restructuring in events like that
studied by \citet{Schrijver2011b} and \citet{Schrijver2015}.

The flaring rate can be used to investigate the significance of such
triggering, if present. \citet{Dalla2007} searched for differences in
the flaring rate of new active regions depending upon whether each
region was ``paired'' --- defined as emerging within 12$^\circ$ of a
pre-existing region --- or isolated (farther than 12$^\circ$).  They
found only a weak effect, but did note that when flares in a paired
system did occur, they were more common in the pre-existing region.
Notably, \citet{Dalla2007} did not incorporate any threshold on the
size of the new regions in their analysis.
Based upon a theoretical estimate of the strength of new AR -- PEAR
interaction energy that we outline below, we expected that PEARs
should be more strongly affected an emerging AR that is (i) larger and
(ii) closer to a given AR.
As described below, we checked for evidence of a ``dose-response''
relationship for both of these factors, finding (i) no obvious size
dependence, and (ii) clear indications of proximity dependence.


In this paper, our aim is to quantify the influence of new active
regions on the release of magnetic energy via flares in PEARs.  We
first present a simple theoretical approach to quantify the 
%
%
interaction energy between new and pre-existing flux systems (\S
\ref{subsec:theory}), and then apply it to two well-known examples of
flux emergence to investigate this approach in observed configurations
(\S \ref{subsec:ar10488}).  We then systematically identify new active
regions in NOAA records, and statistically investigate the influence
of these regions on the rate of flaring in pre-existing regions.  In
the process of defining a single emergence time for each region, we
identified a common feature of many emergence events: active regions'
magnetic flux often emerges in two phases, with a slow initial phase
followed by a faster subsequent phase, discussed in \S
\ref{subsec:emergtime}.  Our statistical results, discussed in \S
\ref{subsec:epoch}, suggest that emergence of large active regions
    is associated with a significant increase in the rate
    of large flares in PEARs.  Finally, in \S \ref{sec:conclu}, we
    discuss the significance of our results.


\section{Interaction energy}
\label{sec:interg}

\subsection{Theory}
\label{subsec:theory}

Flares and CMEs are thought to be powered by free magnetic energy,
$U^{\rm free}$, which is the difference between two energies,
\be U^{\rm free} \equiv U^{\rm actual} - U^{\rm pot} ~, \label{eqn:free_e} \ee
where $U^{\rm actual}$ is the energy of the actual magnetic field,
\be U^{\rm actual} = \frac{1}{8 \pi} \int dV (\Bvec \cdot \bvec) ~, 
\label{eqn:uact} \ee
and $U^{\rm pot}$ is the energy of the hypothetical, current-free
magnetic field consistent with the same radial photospheric magnetic
field, known as the potential field $\Bvec^{\rm pot}$.

For any given radial photospheric flux
distribution, the potential field $\Bvec^{\rm pot}$ obeys
\begin{equation}
\nabla\times\Bvec^{\rm pot} = \nabla\cdot\Bvec^{\rm pot} = 0 ~,
\label{eqn:curl_free}
\end{equation}
which implies that $\bvec$ can be expressed as the gradient of a
scalar potential,
\begin{equation}
\Bvec^{\rm pot} = -\nabla\chi ~. 
\label{eqn:grad_chi}
\end{equation}
The divergence-free condition on $\bvec$ then implies
\begin{equation}
\nabla \cdot \Bvec^{\rm pot} = \nabla^2\chi = 0 ~,
\label{eqn:laplac}
\end{equation}
i.e., that $\chi$ is a solution to Laplace's equation; as such, the
potential field is uniquely defined for any specified Neumann or
Dirichlet boundary condition.

The potential field can be shown to be the lowest-energy configuration
of coronal magnetic field consistent with $B_r\vert_{r=r_\odot}$
(e.g., \citealt{Priest2014}) . The energy of such configuration is
\begin{eqnarray}
U^{\rm pot} 
&=& \frac{1}{8\pi}\int dV \, (\Bvec^{\rm pot})^2 \label{eqn:potergB} \\
&=& \frac{-1}{8\pi} \int_S dA\, \chi \, \frac{\partial \chi}{\partial r} \\
  \label{eqn:poterg_partialchi}
&=& \frac{1}{8\pi} \int_S dA\, \chi \, B_r 
  ~, \label{eqn:poterg}
\end{eqnarray}
where we made use of the divergence theorem and equation
(\ref{eqn:laplac}), $\hatr$ is the unit normal vector (positive
toward the corona) at the photosphere, and the integral range $S$ is the whole
photosphere. 
(The outer surface is assumed to make no contribution.)
Because $\chi$ and $\Bvec^{\rm pot}$ are determined solely by
$B_r\vert_{r=r_\odot}$, the potential field energy, $U^{\rm pot}$, is a
functional of $B_r\vert_{r=r_\odot}$.
(Note that any constant added to $\chi$ in equation [\ref{eqn:poterg}]
will not change the result, assuming radial flux balance.) 

\citet{Schrijver2003} give a specific method to solve for a version of
this hypothetical, current-free field in spherical coordinates, with
the radial field at the photosphere,
$B_r(r,\varphi,\theta)\vert_{r=r_{\odot}}$, used as a Neumann boundary
condition, and a ``source surface'' at $r_{SS} = 2.5 r_\odot$, where
$\bvec$ is purely radial, as an outer, Dirichlet condition.  As
described by \citet{Altschuler1969}, this outer boundary condition is
meant to reproduce the effect of the outward expansion of the solar
wind on the magnetic field's structure near $r_{SS}$: as the field's
strength falls off with distance above the Sun's surface, the outflow
essentially overcomes magnetic tension to force the field to become
purely radial.
Such a field is known as a potential-field source-surface (PFSS)
model.
The radial-field condition at $r_{SS}$ implies $\chi$ is constant over
this surface; the lack of magnetic monopoles then implies no
contribution from the surface integral at $r_{SS}$ in equation
(\ref{eqn:poterg}), consistent with our assumption there.
Flux systems that extend beyond the source surface are referred to as open
fields; flux systems that do not reach the source surface are referred to as
closed, and typically contain arch-shaped field lines.
Although much of the formalism that we develop here is valid for
potential fields that obey different boundary conditions than the PFSS
solution does, we focus primarily on PFSS models in what follows.
This potential field is of interest because it corresponds to the
coronal field's fully-relaxed state.

The free energy in the coronal magnetic field therefore arises from
the presence of coronal electric currents --- departures from the
current-free, minimum-energy state. Such currents are inferred from
observations of ``non-potential'' magnetic structures, such as
field-aligned H-$\alpha$ fibrils in filaments whose orientations are
inconsistent with potential-field orientations (e.g.,
\citealt{Martin1998}), above-limb rings of enhanced linear
polarization consistent with flux ropes viewed in cross section (e.g.,
\citealt{Dove2011}), radial electric currents at the photosphere,
$J_r$ (i.e., currents normal to the photosphere; e.g.,
\citealt{Leka1996}), and magnetic shear \citep{Leka2003a} along
inversion lines of the normal magnetic field (e.g.,
\citealt{Georgoulis2012, Torok2014}).

How does free energy change?  The coronal magnetic field
evolves in response to changes in the photospheric magnetic field, and
this photospheric driving can inject free energy into the corona. But
the high temperatures and long length scales present in the coronal
magnetic field mean that the coronal conductivity is high, and the
magnetic diffusion time is long. Consequently, the evolution of the
coronal field can usually be taken to be ideal, and the connectivity
of magnetic field lines in the corona can remain unchanged over time
scales very long compared to those of the underlying plasma (an
Alfv\'en crossing time or less). (Notable exceptions to this
constraint occur during flares and CMEs, when local enhancements of
diffusivity allow fast magnetic reconnection to occur [see, e.g.,
  Priest \& Forbes 2000].)  Generally, this constrained evolution
prevents the coronal field, $\bvec$, from relaxing to the lowest
possible energy state, $\Bvec^{\rm pot}$, and leads to the storage of
free magnetic energy. The coronal field, however, should still evolve
toward $\Bvec^{\rm pot}$, even if it may not actually attain this
state.

What can changes in potential fields tell us about free energy in the
actual field?  On its face, this is a strange question, since the
potential field completely neglects the electric currents in the low
corona (by which we mean roughly $r_\odot < r < r_{SS}$) that store
free magnetic energy.  In fact, since the potential field in the
corona only depends upon the radial magnetic field at the photosphere,
significant changes in the PFSS model field might have nothing to do
with evolution of the actual magnetic field in the corona.
The PFSS does, however, approximate the minimum energy state
attainable by the coronal field (which is not necessarily radial at
$r_{SS}$), given the imposed radial field boundary conditions at
$r_\odot$.  Consequently, if a PFSS minimum-energy state changes, for
example, to one with substantially more open flux in a given region,
it is probable that the actual coronal field will evolve to open flux
in that region --- perhaps by magnetic reconnection between open and
closed fields, manifested as a jet or surge, or by ejecting flux into
interplanetary space, manifested as a CME. (It is likely that, in
order for opening the field in a region to be energetically favorable,
field elsewhere must close, necessitating magnetic reconnection
[\citealt{Antiochos1999a}].)  This idea is supported by a study by
\citet{Schrijver2005}, who compared PFSS model fields with actual
coronal loops in TRACE EUV observations, and found that, statistically, AR
coronae whose magnetic structures were more non-potential were more
likely to flare than coronae that appeared more potential.  In this
framework, then, significant changes in the potential field might
indicate the availability of a new minimum-energy configuration that
might lead to a flare or CME.

Given the expected relationship between non-potentiality and flare/CME
activity, we might learn something useful from changes in global
potential field models due to the emergence of a new active region.
For simplicity, we restrict our attention to the special case in which
the new region is isolated; that is, the photospheric areas surrounding
it lack the spatially coherent fields associated with active regions.  
In order to characterize the effect on pre-existing flux systems
caused by emergence of the new flux system, we consider three
hypothetical magnetic fields: (i) the potential field 
arising from magnetic flux outside of the area in which the new region
emerges, 
which we refer to as the ``background'' potential state (into which
the region emerges); (ii) the potential field that would arise if
magnetic flux from the new region were the only flux present on the
entire photosphere, which we refer to as the ``added'' potential
field; and (iii) the potential field after the new region's emergence,
arising from all flux over the entire photosphere, which we refer to
as the ``combined'' potential state.
%
%
We define the background potential field's photospheric boundary
condition to be $B_r^{\rm bg}$, and specify that the site of
emergence of the new region is exactly field-free in $B_r^{\rm bg}$
(though weak photospheric fields are ubiquitous).
We define the added potential field's boundary condition to be
$B_r^{\rm new}$, and equal to the radial magnetic field of the new
region at a given stage of its emergence; outside of the emerging
region, the radial field is zero.  
By construction, $B_r^{\rm bg}$ is zero where $B_r^{\rm new}$ is
  not, and {\em vice versa}, and we can define the combined potential
field's boundary condition to be a sum of the other two fields'
conditions:
\begin{equation}
B_r^{\rm combo} = B_r^{\rm bg} + B_r^{\rm new}  ~.
\end{equation} 
Essentially, we have disjointly partitioned the photosphere into two
complementary areas: one containing the new region, and one containing
the rest of the photospheric surface $S$.
The potential field in the volume above $r_\odot$ depends linearly upon
the surface field; accordingly, $\Bvec = \Bvec^{\rm bg} + \Bvec^{\rm
  new} $.  Note, however, that the potential energy $U[B_r]$ for a
given $B_r$ does not depend linearly on $B_r$. Defining $U^{\rm new} =
U[B_r^{\rm new}]$ and $U^{\rm bg} = U[B_r^{\rm bg}]$, we notice that
$U^{\rm combo} = U[B_r^{\rm combo}]$ generally does not equal $U^{\rm
  new} + U^{\rm bg}$. Instead, we have
\begin{eqnarray}
U^{\rm combo} 
&=& \frac{1}{8\pi}\int dV (\Bvec^{\rm new} + \Bvec^{\rm bg})^2 \\
&=& U^{\rm new} + U^{\rm bg} + 
\frac{2}{8\pi} \int dV \, \Bvec^{\rm new} \cdot \Bvec^{\rm bg}  \\
&=& U^{\rm new} + U^{\rm bg} + U^{\rm int} ~,
\label{eqn:interaction}
\end{eqnarray}
where we have defined the cross term to be 
\be U^{\rm int} 
\equiv \frac{1}{4 \pi} \int dV (\Bvec^{\rm new} \cdot \Bvec^{\rm bg}) 
~. \label{eqn:uint_def}
\ee
For $r>r_\odot$, $\Bvec^{\rm new}$ and $\Bvec^{\rm bg}$ generally have
overlap, so this term reflects the relation between the newly emerged
active region and any PEAR(s), and we call $U^{\rm int}$ the ``interaction
energy.''
The interaction energy can be computed several ways:
\bea U^{\rm int} &=& U^{\rm combo} - (U^{\rm bg} + U^{\rm new}) \textnormal{, or} 
\label{eqn:uint_difference} \\
&=& \frac{1}{4 \pi} \int_S dA \, \chi^{\rm bg} \, B_r^{\rm new} \textnormal{, or} 
\label{eqn:uint_chi_bg} \\
&=& \frac{1}{4 \pi} \int_S dA \, \chi^{\rm new} \, B_r^{\rm bg}
\label{eqn:uint_chi_new}
~. \eea 
Note that $U^{\rm combo}, U^{\rm bg}$, and $U^{\rm new}$ in equation
(\ref{eqn:uint_difference}) can be determined from just the
post-emergence radial photospheric magnetic field.
Because the potential function, $\chi$, decreases with distance from
its source region(s), the interaction energy can be relatively large when
the new active region emerges very near some PEARs.
We also note that the interaction energy scales with the size of the
new region: a new region with more flux will, in the generic case,
produce a larger interaction energy.

What is the physical significance of the decomposition of $U^{\rm
  combo}$ in equation (\ref{eqn:interaction})?  We observe that
$U^{\rm bg}$ is a lower bound on the energy prior to the new region's
emergence, and $U^{\rm new}$ is a lower bound on the energy added by
this emergence.  Hence, $U^{\rm limit} \equiv U^{\rm bg} + U^{\rm new}$ is
a lower limit on the energy of the actual post-emergence coronal
field: at least this much magnetic energy must be present in the
corona.  This lower limit on the energy actually present is distinct
from the lowest possible energy of the post-emergence field (i.e., the
post-emergence potential state), which is given by $U^{\rm combo}$.
In cases when $U^{\rm int} < 0$, i.e., when $\Bvec^{\rm new}$ and
$\Bvec^{\rm bg}$ are oppositely directed in a volume-averaged sense,
then $U^{\rm limit} > U^{\rm combo}$.  For such cases, we know the
magnetic free energy is at least as large as the difference between
these two energies,
\be U^{\rm free} > U^{\rm limit} - U^{\rm combo} = - U^{\rm int}
\textnormal{, when } U^{\rm int} < 0 ~. 
\label{eqn:free_energy_bound} \ee
Heuristically, $\Bvec^{\rm new}$ and $\Bvec^{\rm bg}$ being oppositely
directed in some region can indicate that magnetic reconnection
between the new and background flux systems in the actual field is
energetically favorable there.
(In the post-emergence potential field, $\Bvec^{\rm new}$ and
$\Bvec^{\rm bg}$ are superposed; the existence of a null point is
plausible in regions where they are oppositely directed.)
To the extent that magnetic reconnection is associated with flares and
CMEs, we therefore expect that active region emergence events with
$U^{\rm int} < 0$ exhibit a greater likelihood of flare and / or CME
activity.
This proposition could be tested by (i) computing $U^{\rm int}$
  for a statistically significant sample of emergence events and (ii)
  characterizing associated flare and CME rates. We lacked time to
  pursue such efforts as part of this study.

We expect that emergence situations with $U^{\rm int} < 0$ should be
relatively common.
Hale's law implies that each magnetic polarity in a new active region
will tend to lie closer to opposite polarities in any same-hemisphere
PEARs than to like polarities in those PEARs.
Similarly, Hale's law also implies that when a new region emerges at
the same longitude as a pre-existing region in the opposite
hemisphere, their leading polarities will tend to be favorably
oriented for trans-equatorial reconnection, as will their following
polarities.
This suggests that active region emergence will often create
configurations with $U^{\rm int} < 0$.


%

We emphasize that the free magnetic energy represented by the negative
interaction energy introduced by emergence of a new active region is
independent of whatever free magnetic energy is contained within that
active region's fields.
The interaction energy depends solely upon the distribution of magnetic
flux of both the emerging region and the global magnetic environment
into which it emerges.
Consequently, even the emergence of an active region that had no
internal free energy --- i.e., one that contained no internal electric
currents --- could, in principle, introduce free energy into the
corona.
(Free energy requires the existence of currents, and in this case,
currents would flow on the separatrix between the new and pre-existing
flux systems.)
Qualitatively, the origin of this energy can be understood in topological
terms: it arises from the significant differences in magnetic
connections that will generally be present between the actual and
potential post-emergence magnetic fields, at least until reconnection
between new and pre-existing flux systems occurs (e.g.,
\citealt{Longcope2005a, Tarr2012, Tarr2014}).
We therefore refer to free energy arising from the emergence of a new
region as ``topological'' free energy, as opposed to the ``internal''
free energy associated with currents present in the emerging flux
system (e.g., \citealt{McClymont1989, Leka1996}) and pre-existing
fields.

\subsection{Case studies of AR10488 and AR11158}
\label{subsec:ar10488}

To demonstrate computation of the interaction energy and investigate
its usefulness, we present two case studies of well-known examples of
emerging active regions, AR 10488 and AR 11158.  
The former, a relatively large region, emerged near the pre-existing
AR 10486, which was one of the largest and most flare-productive
regions of the previous solar cycle (see, e.g.,
\citealt{Kazachenko2010}).
The latter, also a large region, was the first large region to emerge
on the central disk after the launch of the Helioseismic and Magnetic
Imager (HMI; \citealt{Scherrer2012, Schou2012}), and produced an X2.2
flare (see, e.g., \citealt{Sun2012}).  It also emerged near a
pre-existing region, AR 11156.  \citet{Chintzoglou2013} noted that AR
11158 emerged in two phases (see, e.g., their Figure 3), with an initial
``front'' of ``somewhat weaker and fragmented'' flux followed by a
``surge'' of stronger flux that produced rapid growth of flux in the
region.

For our purposes, we want to estimate energies for various boundary
conditions via equation (\ref{eqn:poterg}), so we need global
photospheric fields.  Since the photospheric field is not directly
observed over the entire solar surface, we must use either out-of-date
observations or model results (based upon out-of-date observations)
for far-side fields.  We opt to use global photospheric fields
generated by the LMSAL forecast model (\citealt{Schrijver2003}; and
available online\footnote{
http://www.lmsal.com/solarsoft/archive/ssw/pfss\_links\_v2/
}).

\begin{figure}[htb]
\psfig{figure=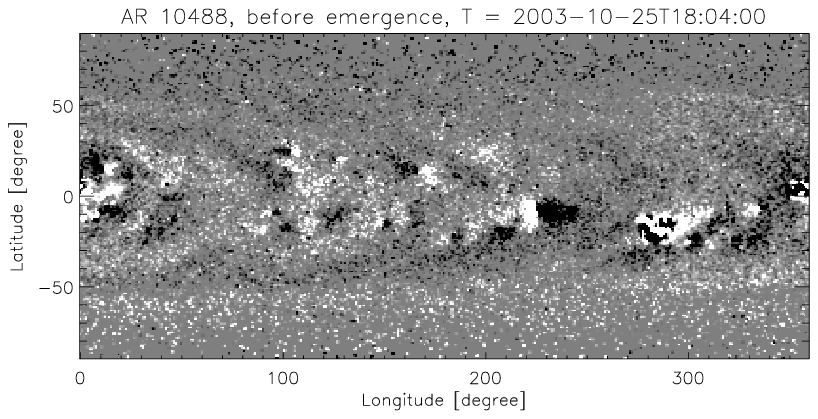, width=4.5in} \\ 
\psfig{figure=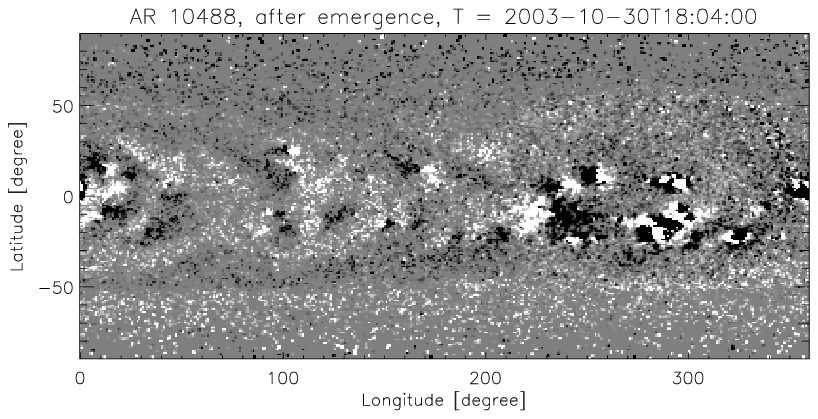, width=4.5in} \\
\epsfig{figure=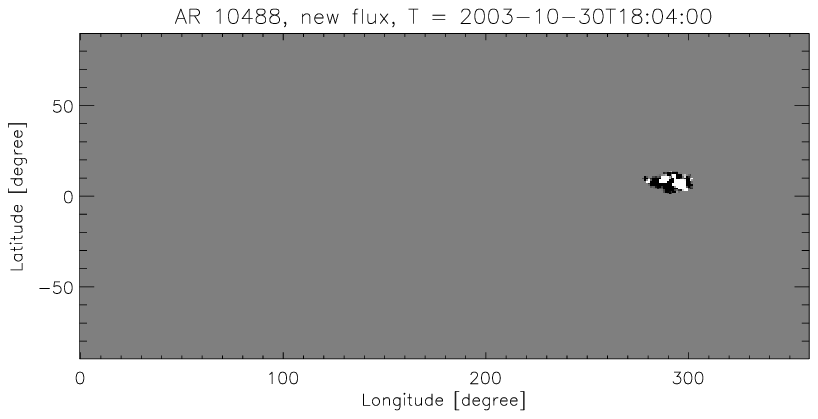, width=4.5in} %
\caption{\footnotesize Surface radial magnetic field maps ---
  pseudo-magnetograms --- from the LMSAL forecast model show the
  pre-emergence global field (top row), the post-emergence global
  field (middle row), and the extracted new flux (bottom row) added to
  the global field by the emergence of AR 10488.  White/black are
  outward/inward radial flux densities, respectively; grayscale saturation is
  set to $\pm 20 \mxcm$.}
\label{fig:magnetograms_10486}
\end{figure}

\begin{figure}[htb]
\epsfig{figure=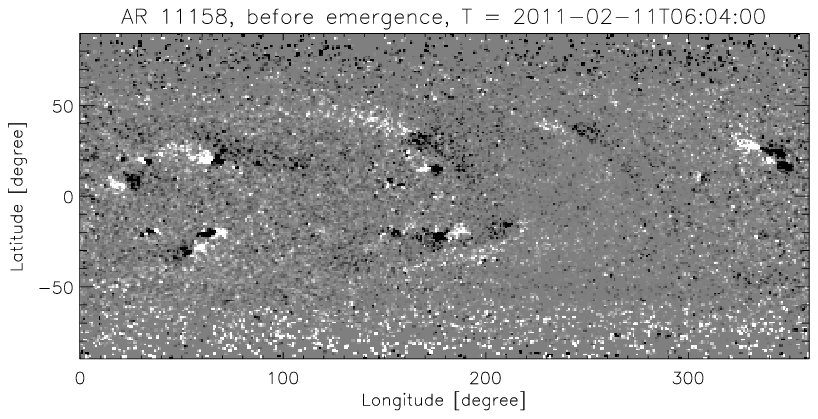, width=4.5in} \\
\epsfig{figure=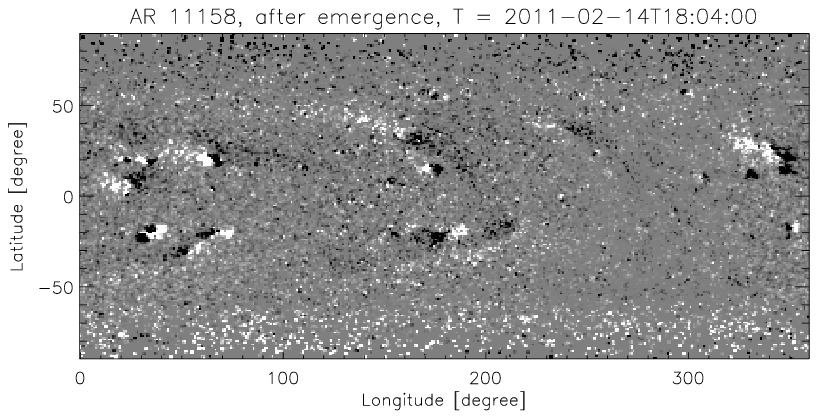, width=4.5in} \\
\epsfig{figure=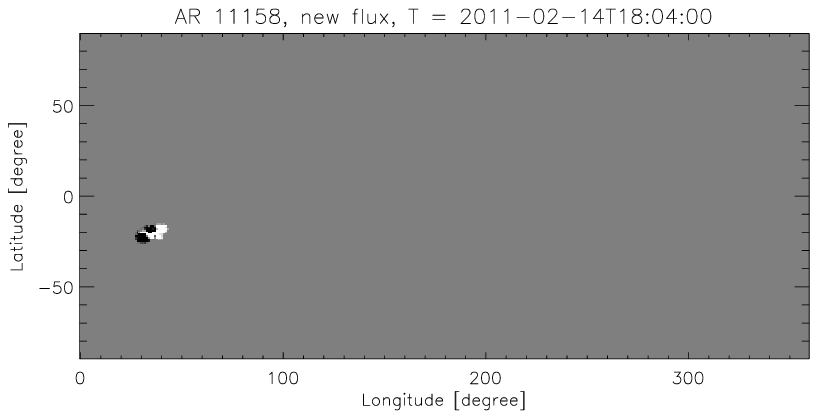, width=4.5in}
\caption{\footnotesize As in previous figure, surface radial magnetic
  field maps showing the pre-emergence global field (top row), the
  post-emergence global field (middle row), and the extracted new flux
  (bottom row), in this case from the emergence of AR 11158.}
\label{fig:magnetograms_11158}
\end{figure}

\begin{figure}[htb]
\centering \epsfig{figure=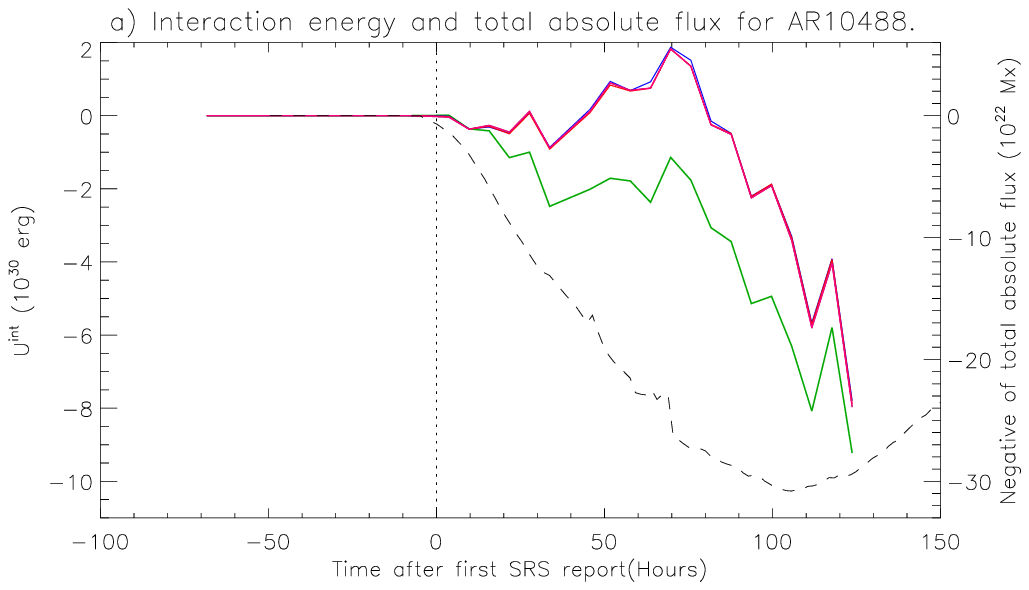, width=4.5in, clip=true}
\psfig{figure=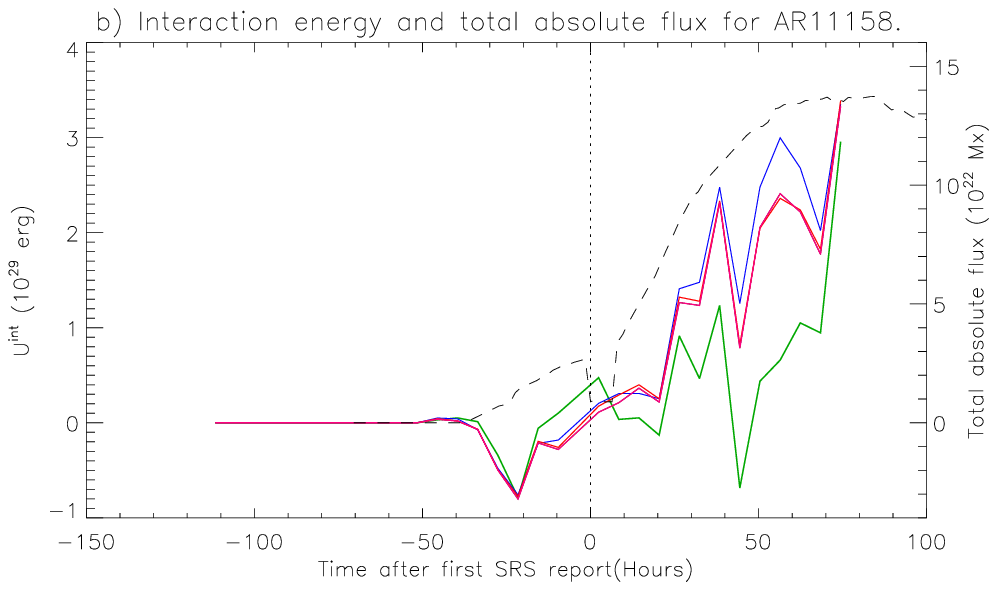, width=4.5in, clip=true}
\caption{\footnotesize Plots of interaction energies (solid lines)
  and total unsigned magnetic fluxes (dashed) for AR 10488 (left) and
  AR 11158 (right). Solid lines represent the result for $U^{\rm int}$
  with maximum $\ell$ values of 32, 64, 128 and 256 in the spherical
  harmonic transforms, and are colored green, blue, pink, and red
  respectively.
  %
  %
  It can be observed that the curves converge with a increasing
  resolution (higher $\ell_{\rm max}$); most nearly overlap in the
  10488 case. The disagreement between the total unsigned flux and
  $U^{\rm int}$ curves implies that $U^{\rm int}$ is not trivially
  related to total flux. }
\label{fig:erg}
\end{figure}

We retrieved surface magnetic field data at the model's six-hour
cadence for 72 hours before and after 00:30 UT on the date of each
region's first appearance in one of the daily Solar Region Summary
(SRS) reports, prepared by USAF and 
NOAA.\footnote{These reports are online; see 
ftp://ftp.swpc.noaa.gov/pub/warehouse/README}
For each magnetic map, we identify the area encompassing the new
region's flux as the contiguous region containing all pixels with flux
densities (i.e., pixel-averaged field strengths) greater than $15
\mxcm$. We then create a ``new-flux'' magnetic field map containing
only the new region's flux, and a ``background'' map, with this flux removed.  
In Figures \ref{fig:magnetograms_10486} and
  \ref{fig:magnetograms_11158}, we show pre-emergence,
    post-emergence, and new-flux magnetic field maps.
For each map, scalar potentials were computed from the radial magnetic
fields
via the IDL spherical harmonic transform software released with
the PFSS package in SolarSoft \citep{Freeland1998}.
We then computed $U^{\rm int}$ via equation
(\ref{eqn:uint_difference}), by separately using the scalar potentials
and radial field for each map in equation (\ref{eqn:poterg}) for each
energy term.
For each stage of the emergence, we computed the interaction energy.
We also varied the maximum $\ell$ value in the summations over
spherical harmonics $Y_{\ell m}$ for each stage, with $\ell_{\rm max}
\in \{32, 64, 128, 256\}$.  Our energy estimates converged by
$\ell_{\rm max} = 128$ (Figure \ref{fig:erg}).

Figure \ref{fig:erg} shows the time variation of total unsigned flux
and interaction energy $U^{\rm int}$ as the emergence proceeds in each
region.  One can see that $U^{\rm int}$ does not depend trivially on
the total unsigned flux: while both quantities tend to grow with the
emergence, they do not agree closely.
From this, we conclude that the interaction energy depends on magnetic
structure that is not simply related to the total unsigned magnetic
flux.

The behavior of $U^{\rm int}$ in the regions differs: it swings
strongly negative in AR 10488, but stays mostly positive in AR 11158.
While both AR 10488 and AR 11158 produced large flares, we are
interested in flare activity in their neighbors.  Given AR 10488's
negative interaction energy, the presence of global free energy due to
this emergence is indicated.  Its nearest neighbor, AR 10486, produced
several large flares; we do not know, however, whether those flares
were triggered by the emergence of AR 10488.  
In addition, we faced a practical problem when computing the
interaction energies due to the emergence of AR 10488: magnetic flux
from far-side PEARs was rotating around the Sun's east limb in the
interval over which we computed interaction energies, and was steadily
incorporated into the potential field computations.  These other
``new'' active regions confound unambiguous attribution of the
negative interaction energy to the emergence of AR 10488.  
(Given the lack of ``$4\pi$'' observations of the Sun's photospheric
magnetic field, all coronal field models will unavoidably suffer from
similar deficiencies.)
In contrast, AR 11158 did not exhibit a negative interaction energy;
and its nearest neighbor, AR 11156, did not produce any significant
flares during the interval that we studied (only two B-class flares).
While the behavior of the neighbors of ARs 10488 and 11158 is
consistent with our expectations based upon their interaction
energies, a statistical investigation of interaction energies and
simultaneous flare activity would be needed to draw any conclusions
about whether interaction energies are significantly associated with
flaring.

We note, however, that the values we find for $U^{\rm int}$ in both
cases (of order 10$^{29}$ - 10$^{30}$) are relatively small compared to the
energy released in large flares and CMEs, which often exceed 10$^{32}$
erg (e.g., \citealt{Emslie2012}).
Both of these cases involved relatively large active regions emerging
near PEARs, so the interaction energies we find are probably large
compared what would be found for more typical emerging regions.
Given that the values of the interaction energies we find are far below
the estimated energies released in relatively large flares, it is
possible that the interaction energies computed in our approach do
not quantify any significant physical property associated with flare
activity.
We expect that global, time-dependent models of non-potential
structures in coronal magnetic fields (e.g., \citealt{Yeates2008}, or
a larger-scale version of the model developed by \citealt{Cheung2012})
could more accurately capture the energetic coupling between active
regions, albeit at considerably greater computational expense than our
approach.

\section{Emerging Regions \& Flaring Rates}
\label{sec:flrrat}

Despite of the shortcomings in our interaction-energy approach for
estimating the energetic consequences of new-region emergence, we
still expect that the emergence of new regions can create additional
topological free magnetic energy in the corona.  As noted above, this
is because the global magnetic connections in the actual
post-emergence magnetic field will initially differ significantly from
the potential post-emergence magnetic field.
This change in free energy introduced by new emerging region could
result in an increase of flares in PEARs.

\subsection{Estimating emergence times}
\label{subsec:emergtime}

To associate flares in PEARs with a new active region's emergence, it
is first necessary to specify an emergence time for the new
region. For this purpose, we developed two different sets of
new-region emergence times, which trade off accuracy for the number of
available cases.

The first and most direct source of estimated emergence times is the
SRS records. We have such records from 1996 to the present, typically
produced at at 00:30UT every day. The date of the earliest report of
each active region then gives a rough estimate for its emergence
time. As will be shown, this often agrees to within 24 hours of a more
precisely determined emergence time.  Using the SRS emergence time has
two disadvantages. First, the record for each active region starts
when human observers recognize the new flux system as an active
region, so it is typically delayed by one day or two after the first
appearance of new flux \citep{Leka2013}, and the threshold for
recognition of new flux as an active region is unclear.
Further, as noted by \citet{Dalla2008}, there is a detection bias in
this catalog: new regions that emerge toward the limbs are often not
immediately detected (by either human observers or algorithms, e.g.,
\citealt{Watson2009}), an issue we revisit briefly below.
Second, the inherent time resolution is limited, because SRS reports
are made only once every 24 hours.

Another way to get a more accurate emergence time for a new AR, with
physically meaningful criteria, is to analyze the growth of the
region's magnetic flux in full-disk magnetograms of the line-of-sight 
field recorded by the Michelson-Doppler Interferometer (MDI)
\citep{Scherrer1995}.  These magnetograms have 2'' , pixels, and
were nominally recorded at 96-minute cadence.

To analyze a given emergence, we must identify all pixels with
associated with a new region as it emerges.
For a sample of 186 emergence events near disk center (within
45$^\circ$), we developed an automated procedure to do so.
Starting from a region's first appearance in an SRS report, we extract
a sequence of magnetograms from 72 hours before to 72 hours after the
SRS appearance, under the assumption that flux might have been present
before the new region was recognized by the observers that compile the
SRS reports.
We then estimate the region's location at each time in the sequence,
based upon its position in the initial SRS report.  
We smooth each magnetogram with a 6-pixel wide, square spatial boxcar,
and then attempt to determine which magnetic flux in the neighborhood
of the new region's location ``belongs'' to the new region --- versus
being simply noise, or belonging to a PEAR.
We do so by identifying all groups of contiguous, above-threshold
pixels, using the label\_region algorithm native to IDL.
To exclude noise when relatively little flux has emerged, we only
group pixels with absolute LOS magnetic field, $|B_{\rm LOS}|,$ above
a more restrictive threshold $40 \mxcm$ (cf., $15 \mxcm$ used
previously, for mature active region fields in the lower-resolution
LMSAL forecast model).
We then associate groups of pixels that are among both (i) the three
closest and (ii) the three largest with the new region. 
To avoid including flux from other active regions in this patch, we
exclude areas closer to the time-adjusted SRS-reported position of any
PEARs than to the new region's time-adjusted location.
To produce a time series of flux associated with the new region, we
then sum the unsigned LOS flux in all groups of pixels that
belong to the new region.
Since we are primarily concerned with short-timescale, relative
changes in new-region flux, we did not correct pixel areas for
foreshortening.

\begin{figure}[htb]
  \psfig{figure=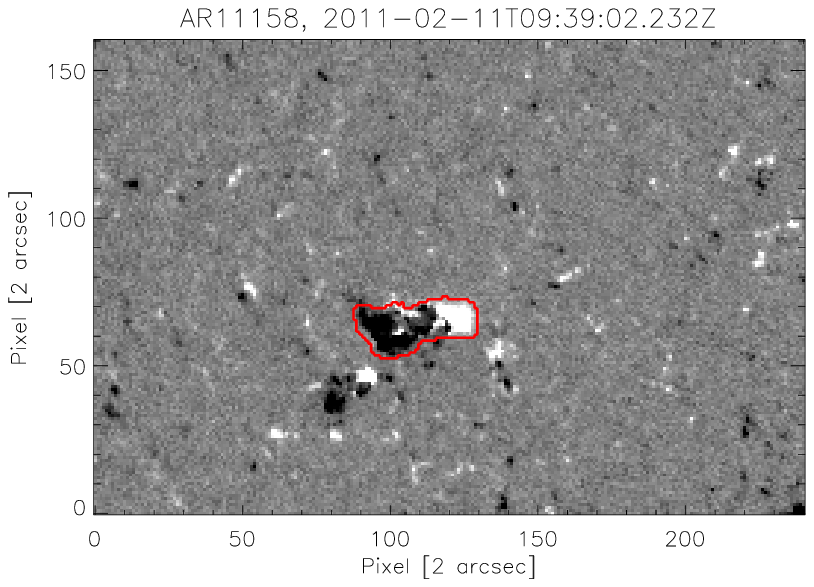, width=2.25in} %
  \psfig{figure=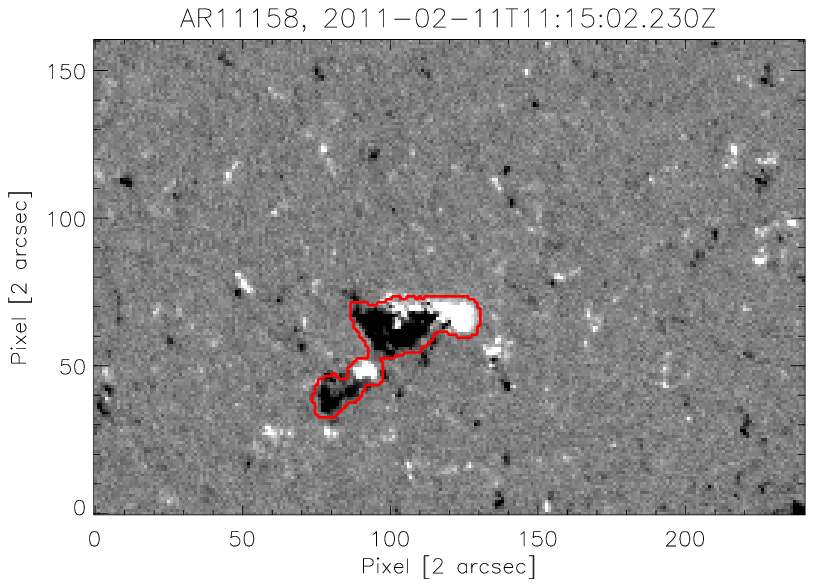, width=2.25in} \\
  \psfig{figure=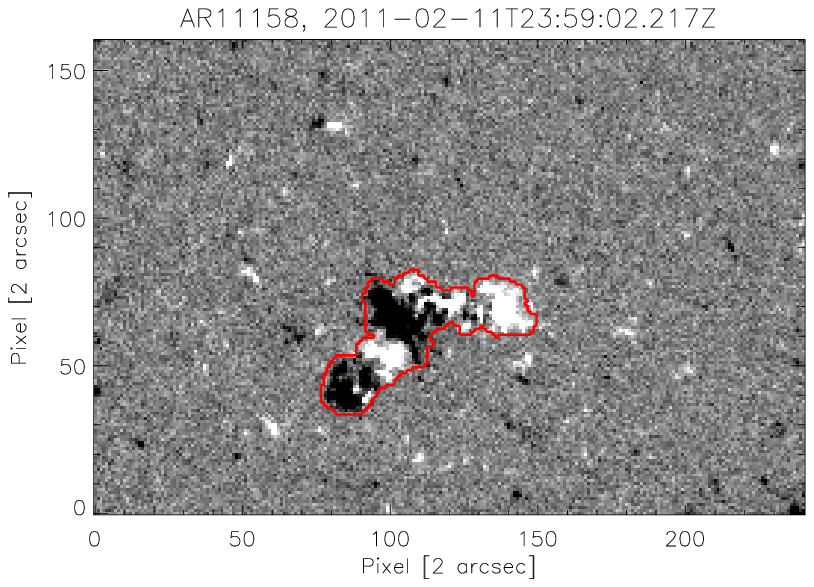, width=2.25in} %
  \psfig{figure=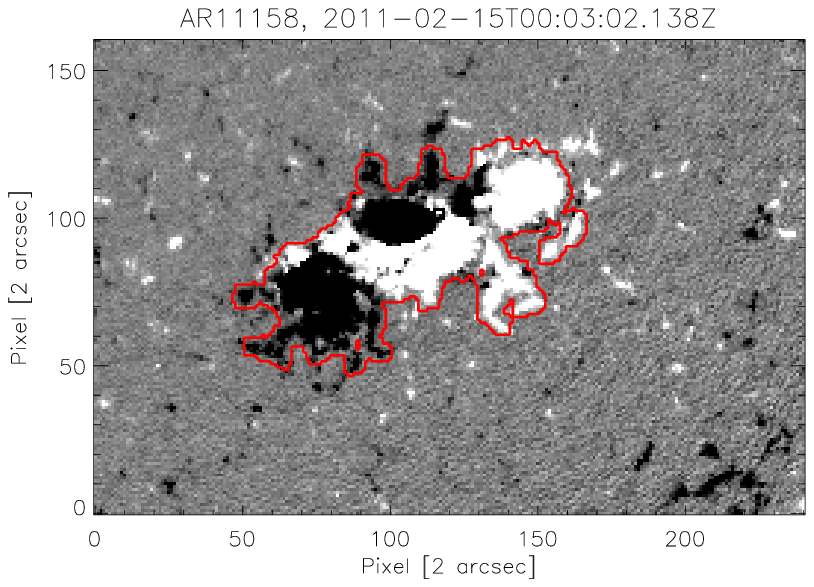, width=2.25in} \\
\caption{\footnotesize Magnetograms of AR 11158, with contours showing
  flux associated with the emerging region.  Between the top-left and
  top-right magnetograms, an initially ungrouped emerging bipole
  becomes grouped with the region, as the emerging bipoles grow toward
  each other.  The bottom-right magnetogram shows the region after
  nearly all of its flux has emerged, about 100 minutes before the
  X2.2 flare.}
\label{fig:auto_flux}
\end{figure}

As an example of the method, in Figure \ref{fig:auto_flux}, we show
four MDI magnetograms of AR 11158, with contours outlining flux associated
with the new region.
The first three magnetograms show the growth of the region during its
initial emergence phase, on 2011-02-11. The final magnetogram shows
the region at the start of 2011-02-15, after most of its flux has
emerged, about 100 minutes before the start of the X2.2 flare.
Between the first and second frames, an initially ungrouped emerging bipole is
subsequently grouped with the new region as the bipoles grow toward
each other.  
As noted above, this region experienced a more rapid phase of
emergence that began around 18:00 UT on 2011-02-12
\citep{Chintzoglou2013}, which is shown in the bottom panel of
Figure \ref{fig:erg}.

\begin{figure}[htb]
\centering
a) \psfig{figure=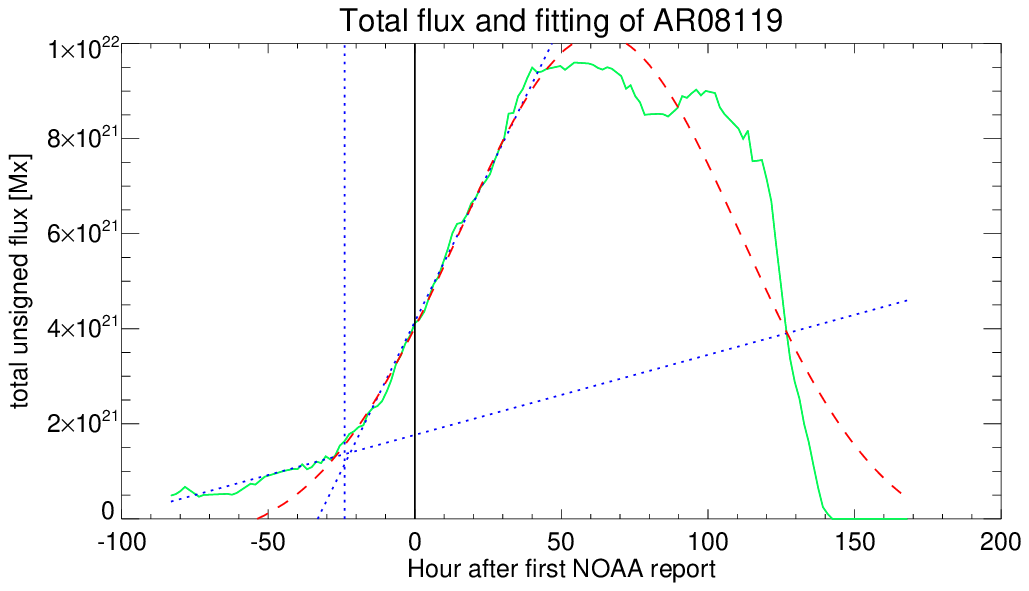, width=3.0in} \\ 
b) \psfig{figure=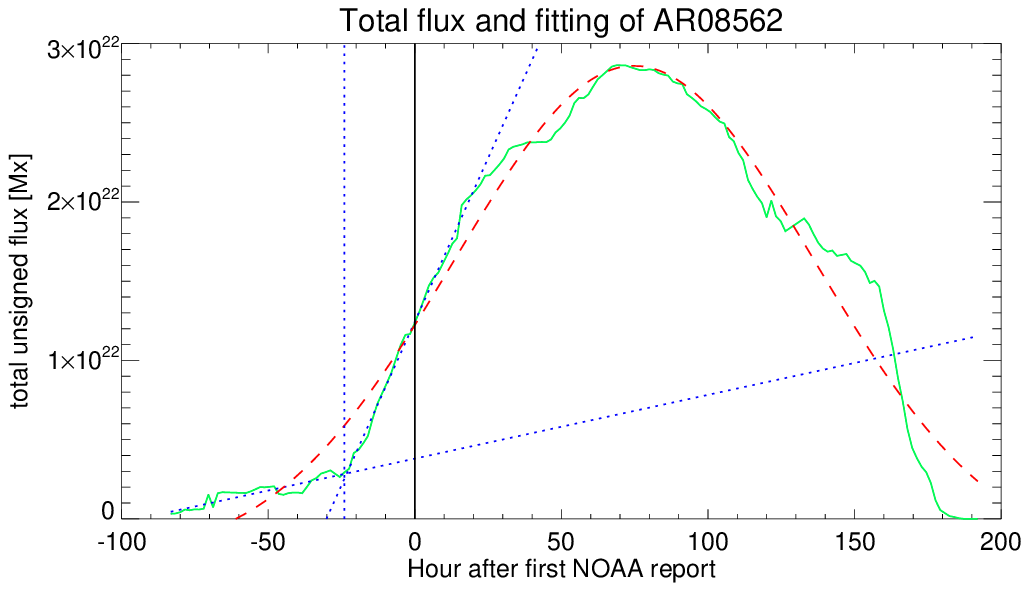, width=3.0in} \\ 
c) \psfig{figure=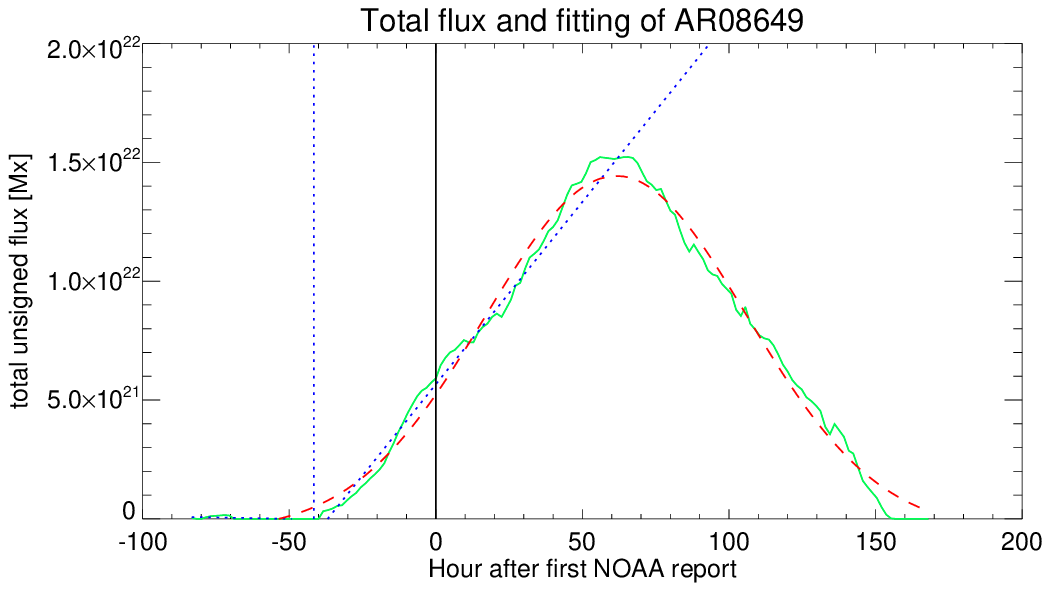, width=3.0in} \\ 
d) \psfig{figure=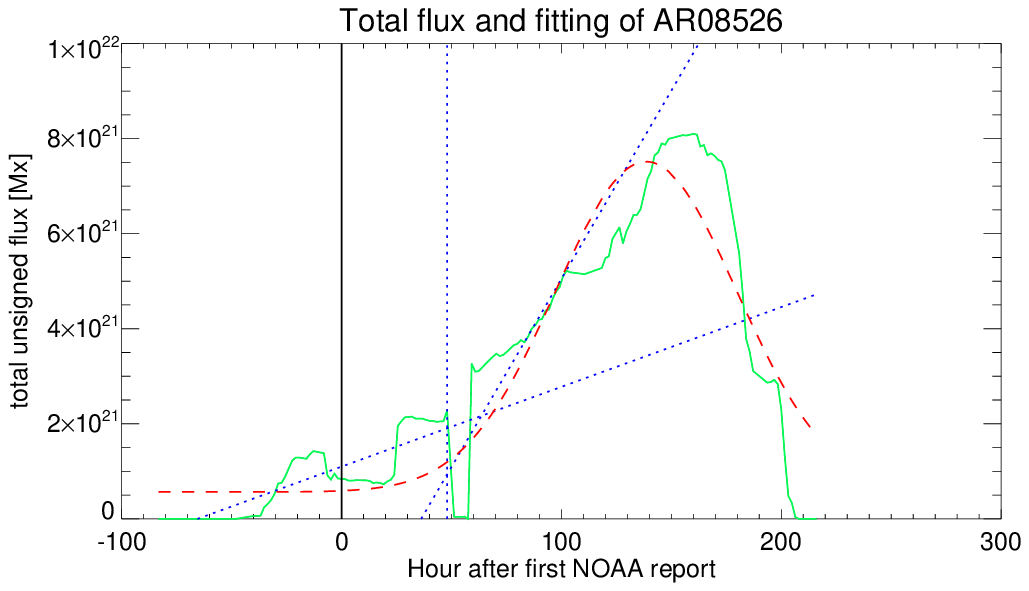, width=3.0in} 
\caption{\footnotesize Examples of curve fitting of total unsigned
  flux to find emergence time.  Green lines in all panels show flux
  versus time in a given emergence region.  Red dashed lines show a
  Gaussian fit to the flux profile. The tilted dotted lines represent
  the two-slope, piece-wise continuous linear fits to the start of the
  flux-versus-time curve, while the vertical dotted line shows the
  emergence time determined from these fits, set by the
  gradual-to-rapid transition in the rate of emergence.  Vertical
  solid lines (always at zero) show the time of each region's first
  SRS report.  Panels a) and b) show examples of regions with
  two-phase emergence; the gradual-to-rapid emergence transition is
  smooth in the former, but more sudden in the latter.  The latter
  behavior was more common in the regions we studied.  Panel c) shows
  a region with a single emergence phase; the fitted, pre-emergence
  slope is zero.  Panel d) shows a case in which our flux
  identification algorithm failed, for a subset of time steps, to
  consistently determine which magnetogram flux ``belongs'' to the new
  AR (versus nearby PEARs, or the background) during the emergence.}
\label{fig:fluxplot}
\end{figure}

In practice, we find that this automated approach can sometimes fail
to properly identify pixels associated with a given new region,
typically for a few consecutive magnetograms.  This can cause
transient dips in flux (e.g., Figure \ref{fig:fluxplot}d).
In most cases, however, it performs well enough:
in our sample of 186 emergence events, we manually rejected 70 for
having unphysical jumps in the flux-versus-time curves due to poor flux
attribution.
Although the method is imperfect, we did not invest further effort to
optimize it, since we only use this tool in an ancillary investigation
of emergence times.



We chose to define the start of a region's emergence to be at the
break point of a two-slope, piece-wise continuous linear fit to its
flux-versus-time curve.
This approach grew out of our observation that in many regions (e.g.,
AR 11158), flux (as identified by our algorithm) emerges in two phases:
an initial, gradual phase, followed by a phase of more rapid emergence
(e.g., Figure \ref{fig:fluxplot}, panels a and b).
The location of the break point between the two linear fits was selected to
minimize the summed squared error in the overall fit.
The start of the fitting interval was 84 hours prior to the first SRS
report containing the new region.  We set the endpoint of the time
interval for our curve fitting
to be the earlier of either (i) the time of the first NOAA record or
(ii) the time at which the region reached 75\% of its maximum unsigned
flux.
%
%
This guarantees that the flux curve to be fitted is almost linear after the
turning point, so the piece-wise linear fit finds the dramatic change
of flux growth well.  We treat this turning point as the emergence
time, since in many cases it captures the start of significant emergence.
We also found many regions in which flux emerges at a single, roughly
constant rate, and this fitting procedure also works well for such
cases; the slope for the pre-emergence interval is simply zero (Figure
\ref{fig:fluxplot}c).

%

\begin{table}
\footnotesize
\caption{For each emerging region that we study, we show the start and
  stop times of our fitting interval, and the two slopes found by our
  algorithm.}
\label{tab:slopes1}
\resizebox{!}{0.4\textheight}{
  \begin{tabular}{rcccc}
\hline
NOAA & SRS Appearance & Fitted Appearance & Init. Slope $a_1$ & Final Slope, $a_2$ \\
AR \# & [YYYY-MM-DD hh:mm] & [YYYY-MM-DD hh:mm]
& [2 $\times 10^{16}$ Mx/s] & [2 $\times 10^{16}$ Mx/s] \\
\hline
 7982 & 1996-08-08 00:00 & 1996-08-08 17:36 &  1.56 &  6.11 \\ 
 8119 & 1997-12-09 00:00 & 1997-12-08 00:03 &  0.97 &  7.18 \\ 
 8169 & 1998-02-28 00:00 & 1998-02-26 11:12 &  0.00 &  3.95 \\ 
 8193 & 1998-04-05 00:00 & 1998-04-04 06:24 &  4.55 & 28.28 \\ 
 8200 & 1998-04-10 00:00 & 1998-04-08 19:12 &  1.04 & 11.62 \\ 
 8203 & 1998-04-15 00:00 & 1998-04-14 11:15 &  0.57 & 17.19 \\ 
 8226 & 1998-05-24 00:00 & 1998-05-23 03:11 &  0.54 & 26.55 \\ 
 8404 & 1998-12-07 00:00 & 1998-12-05 20:48 &  0.70 & 20.13 \\ 
 8506 & 1999-04-04 00:00 & 1999-04-03 11:12 &  2.32 & 17.91 \\ 
 8562 & 1999-06-01 00:00 & 1999-05-31 00:00 &  2.31 & 23.67 \\ 
 8582 & 1999-06-12 00:00 & 1999-06-10 14:24 &  0.25 &  7.85 \\ 
 8649 & 1999-07-28 00:00 & 1999-07-26 06:24 & -0.13 &  8.80 \\ 
 8737 & 1999-10-20 00:00 & 1999-10-18 06:23 &  1.14 & 10.69 \\ 
 8743 & 1999-10-26 00:00 & 1999-10-29 04:48 &  2.00 & 11.22 \\ 
 8757 & 1999-11-06 00:00 & 1999-11-05 11:12 &  0.55 & 10.31 \\ 
 8797 & 1999-12-14 00:00 & 1999-12-12 11:11 &  0.74 &  7.04 \\ 
 8837 & 2000-01-19 00:00 & 2000-01-17 07:59 &  0.00 & 13.51 \\
 8844 & 2000-01-25 00:00 & 2000-01-24 11:11 &  0.72 & 21.53 \\ 
 8897 & 2000-03-02 00:00 & 2000-03-02 03:15 &  0.66 &  6.31 \\ 
 8904 & 2000-03-09 00:00 & 2000-03-09 03:11 &  3.20 & 18.24 \\ 
 8917 & 2000-03-19 00:00 & 2000-03-18 04:47 &  0.00 & 17.03 \\ 
 8918 & 2000-03-20 00:00 & 2000-03-19 03:11 &  1.41 & 13.83 \\ 
 8926 & 2000-03-24 00:00 & 2000-03-22 19:15 &  0.23 & 12.71 \\ 
 8935 & 2000-03-30 00:00 & 2000-03-27 22:23 &  0.09 &  7.42 \\ 
 8968 & 2000-04-20 00:00 & 2000-04-19 11:15 &  0.99 & 15.00 \\ 
 9135 & 2000-08-16 00:00 & 2000-08-13 17:35 &  1.84 & 11.99 \\ 
 9136 & 2000-08-18 00:00 & 2000-08-16 11:12 & -0.25 & 16.51 \\ 
 9139 & 2000-08-21 00:00 & 2000-08-19 14:23 &  0.84 & 14.21 \\ 
 9144 & 2000-08-26 00:00 & 2000-08-25 04:47 &  0.26 & 24.74 \\ 
 9170 & 2000-09-22 00:00 & 2000-09-20 01:35 &  0.00 & 10.21 \\ 
 9290 & 2000-12-28 00:00 & 2000-12-27 19:11 &  3.18 & 12.22 \\ 
 9291 & 2000-12-30 00:00 & 2000-12-28 15:59 &  0.06 &  8.84 \\ 
 9324 & 2001-01-24 00:00 & 2001-01-21 16:03 &  0.00 &  2.54 \\ 
 9366 & 2001-03-03 00:00 & 2001-02-28 23:59 &  0.48 &  8.02 \\ 
 9408 & 2001-03-29 00:00 & 2001-03-27 08:03 &  0.07 & 12.38 \\ 
 9413 & 2001-04-02 00:00 & 2001-03-31 22:24 &  0.13 &  7.45 \\ 
 9426 & 2001-04-12 00:00 & 2001-04-10 19:12 &  1.25 & 15.31 \\ 
 9432 & 2001-04-19 00:00 & 2001-04-17 11:12 &  1.12 &  9.72 \\ 
 9447 & 2001-05-01 00:00 & 2001-05-01 04:48 &  2.79 & 13.57 \\ 
 9455 & 2001-05-12 00:00 & 2001-05-10 11:12 &  0.35 & 18.01 \\ 
 9484 & 2001-06-02 00:00 & 2001-06-01 16:00 &  0.63 & 17.62 \\ 
 9512 & 2001-06-22 00:00 & 2001-06-21 12:51 &  3.55 & 17.56 \\ 
 9531 & 2001-07-09 00:00 & 2001-07-07 14:24 &  0.25 & 21.83 \\ 
 9553 & 2001-07-24 00:00 & 2001-07-23 04:51 &  0.75 & 10.76 \\ 
 9574 & 2001-08-11 00:00 & 2001-08-10 07:59 &  1.42 & 40.42 \\ 
 9611 & 2001-09-08 00:00 & 2001-09-07 08:03 &  0.37 &  9.35 \\ 
 9631 & 2001-09-21 00:00 & 2001-09-19 16:03 &  0.70 & 13.71 \\ 
 9635 & 2001-09-24 00:00 & 2001-09-22 20:48 &  0.22 &  5.94 \\ 
 9639 & 2001-09-27 00:00 & 2001-09-26 16:03 &  1.95 &  7.40 \\ 
 9645 & 2001-10-02 00:00 & 2001-09-30 04:47 & -0.14 &  7.90 \\ 
 9674 & 2001-10-20 00:00 & 2001-10-19 12:51 &  1.55 & 34.59 \\ 
 9692 & 2001-11-08 00:00 & 2001-11-07 06:23 &  1.84 & 26.65 
\end{tabular}}
\end{table}

\begin{table}
\footnotesize
\caption{Continuation of Table 1.}
\label{tab:slopes2}
\resizebox{!}{0.4\textheight}{
\begin{tabular}{rcccc}
\hline
NOAA & SRS Appearance & Fitted Appearance & Init. Slope $a_1$ & Final Slope, $a_2$ \\
AR \# & [YYYY-MM-DD hh:mm] & [YYYY-MM-DD hh:mm]
& [2 $\times 10^{16}$] & [2 $\times 10^{16}$] \\
\hline
9719 & 2001-11-29 00:00 & 2001-11-26 15:59 &  1.92 &  0.38 \\ 
 9739 & 2001-12-14 00:00 & 2001-12-13 11:15 & -0.12 & 14.79 \\ 
 9748 & 2001-12-21 00:00 & 2001-12-20 14:23 &  0.50 & 15.29 \\ 
 9764 & 2001-12-29 00:00 & 2001-12-29 22:23 &  2.14 & 12.51 \\ 
 9768 & 2002-01-02 00:00 & 2002-01-03 17:39 &  3.56 & 11.71 \\ 
 9786 & 2002-01-17 00:00 & 2002-01-16 01:39 &  0.83 &  4.05 \\ 
 9844 & 2002-02-24 00:00 & 2002-02-23 03:12 &  1.21 &  8.82 \\ 
 9846 & 2002-02-25 00:00 & 2002-02-24 08:00 & -0.06 & 12.72 \\ 
 9868 & 2002-03-12 00:00 & 2002-03-10 14:24 &  1.64 &  7.67 \\ 
 9872 & 2002-03-16 00:00 & 2002-03-14 16:00 &  0.78 &  7.49 \\ 
 9873 & 2002-03-18 00:00 & 2002-03-16 04:48 &  3.51 & 15.43 \\ 
 9912 & 2002-04-19 00:00 & 2002-04-18 03:12 &  1.47 & 14.55 \\ 
 9926 & 2002-04-28 00:00 & 2002-04-25 03:11 & 23.27 &  5.99 \\ 
 9931 & 2002-05-02 00:00 & 2002-04-30 06:23 &  0.02 &  6.85 \\ 
10045 & 2002-07-25 00:00 & 2002-07-25 03:15 &  1.74 & 17.88 \\ 
10050 & 2002-07-27 00:00 & 2002-07-26 22:23 &  1.77 & 30.38 \\ 
10072 & 2002-08-12 00:00 & 2002-08-11 06:27 &  0.33 &  6.73 \\ 
10097 & 2002-09-01 00:00 & 2002-08-30 15:59 &  1.00 &  6.06 \\ 
10099 & 2002-09-02 00:00 & 2002-09-01 01:35 &  0.35 &  8.12 \\ 
10109 & 2002-09-11 00:00 & 2002-09-09 17:35 &  1.26 &  6.74 \\ 
10110 & 2002-09-11 00:00 & 2002-09-09 09:35 &  1.03 &  6.88 \\ 
10132 & 2002-09-23 00:00 & 2002-09-21 09:36 & -0.14 & 15.09 \\ 
10133 & 2002-09-24 00:00 & 2002-09-22 03:11 &  0.00 &  7.57 \\ 
10178 & 2002-11-01 00:00 & 2002-10-30 19:11 &  0.44 & 11.24 \\ 
10188 & 2002-11-07 00:00 & 2002-11-06 04:51 &  1.22 & 16.76 \\ 
10227 & 2002-12-14 00:00 & 2002-12-13 08:03 &  0.00 &  5.31 \\ 
10249 & 2003-01-08 00:00 & 2003-01-06 16:02 &  0.88 &  7.51 \\ 
10253 & 2003-01-11 00:00 & 2003-01-09 23:59 &  0.00 &  9.02 \\ 
10313 & 2003-03-14 00:00 & 2003-03-12 20:48 &  0.00 & 11.24 \\ 
10350 & 2003-04-30 00:00 & 2003-04-28 09:36 & -0.45 & 66.56 \\ 
10381 & 2003-06-10 00:00 & 2003-06-09 16:02 &  0.95 & 10.29 \\ 
10382 & 2003-06-11 00:00 & 2003-06-09 19:14 &  0.01 &  3.88 \\ 
10388 & 2003-06-20 00:00 & 2003-06-19 07:58 &  2.66 & 16.20 \\ 
10491 & 2003-10-28 00:00 & 2003-10-26 14:23 &  0.00 & 10.43 \\ 
10591 & 2004-04-13 00:00 & 2004-04-11 03:15 &  0.00 &  5.39 \\ 
10601 & 2004-05-01 00:00 & 2004-04-30 01:35 &  0.67 & 15.48 \\ 
10605 & 2004-05-05 00:00 & 2004-05-03 12:47 &  0.62 &  7.67 \\ 
10692 & 2004-10-25 00:00 & 2004-10-23 17:36 &  0.01 & 13.14 \\ 
10747 & 2005-04-01 00:00 & 2005-04-01 19:11 &  1.03 &  8.57 \\ 
10770 & 2005-05-30 00:00 & 2005-05-28 23:59 &  0.13 &  6.85 \\ 
10819 & 2005-11-01 00:00 & 2005-10-30 14:24 &  1.36 &  5.04 \\ 
10862 & 2006-03-19 00:00 & 2006-03-16 16:03 & -0.35 &  5.01 \\ 
10869 & 2006-04-07 00:00 & 2006-04-05 01:35 &  0.00 &  4.63 \\ 
10879 & 2006-05-03 00:00 & 2006-05-01 06:23 &  0.05 &  5.09 \\ 
10885 & 2006-05-21 00:00 & 2006-05-19 03:11 &  0.02 &  4.04 \\ 
10917 & 2006-10-20 00:00 & 2006-10-21 04:51 &  2.06 & 19.52 \\ 
10955 & 2007-05-10 00:00 & 2007-05-08 11:11 &  2.06 &  6.52 \\ 
11005 & 2008-10-12 00:00 & 2008-10-10 16:03 &  0.00 & 10.93 \\ 
11027 & 2009-09-23 00:00 & 2009-09-22 01:39 &  0.27 & 11.39 \\ 
11029 & 2009-10-24 00:00 & 2009-10-22 22:27 &  0.39 &  7.89 \\ 
11061 & 2010-04-06 00:00 & 2010-04-05 09:39 &  1.32 & 19.72 \\ 
11105 & 2010-09-03 00:00 & 2010-09-01 19:15 &  0.14 &  6.54
\end{tabular}}
\end{table}

For the 116 emergence cases with smooth flux evolution, we deemed the
fit results to be poor in 11 cases.
In a further 5 cases, the initial part of the flux-versus-time plot
was inconsistent with new flux emergence, being well above zero
(indicating pre-existing flux) and either (i) decreasing or (ii)
flat.
Of the remaining 100 cases, we manually characterized 46 as being more
consistent with single-phase emergence, and 54 as being more
consistent with two-phase emergence.  
Hence, we find about 50\% of emergence events exhibit two-phase
behavior.  
Examples of single- and two-phase emergence events are shown in Figure
\ref{fig:fluxplot}.

\begin{figure}[htb]
\centerline{\psfig{figure=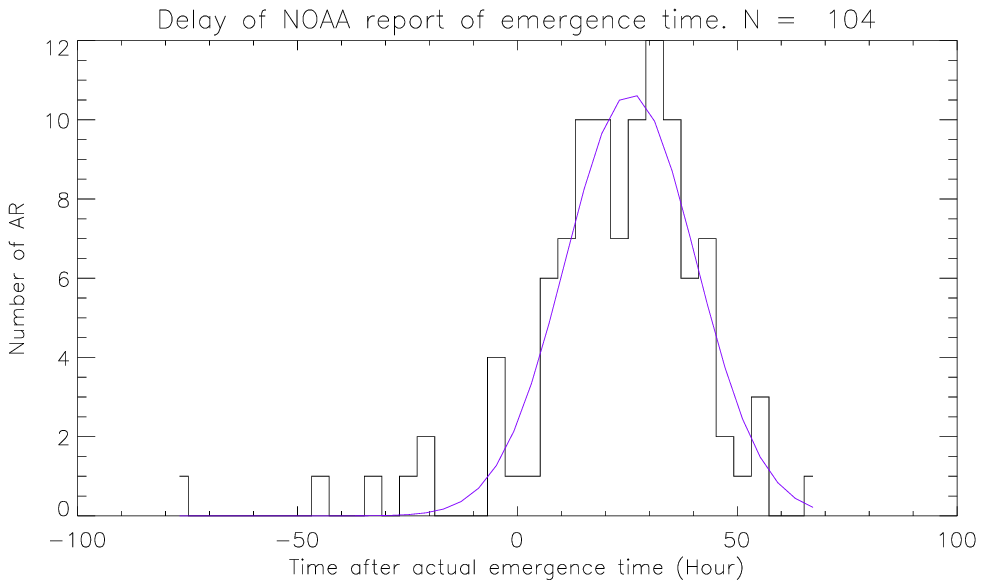, width=4.0in}}
\centerline{\psfig{figure=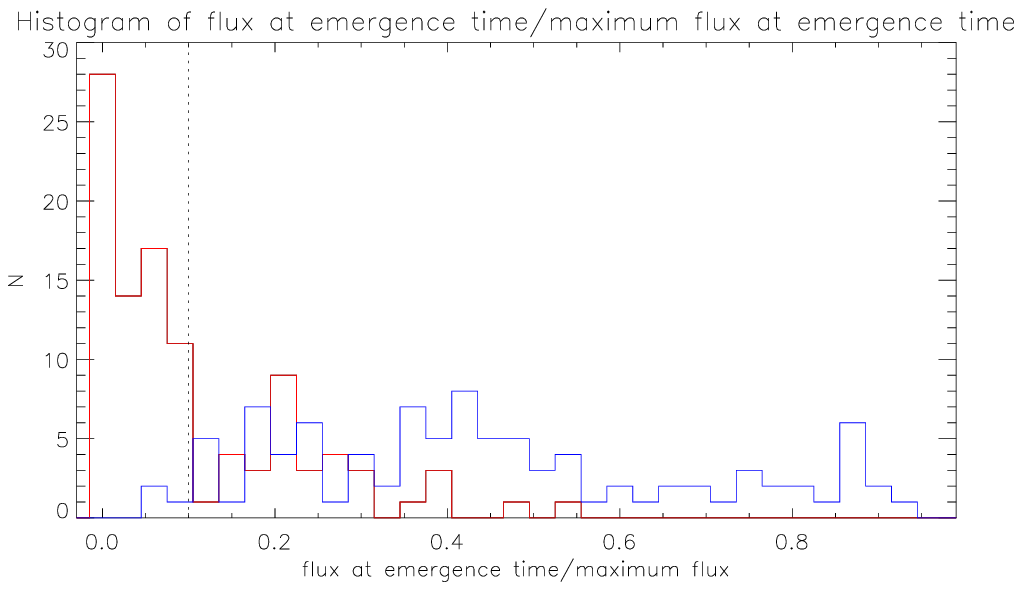, width=4.0in}}
\centerline{\psfig{figure=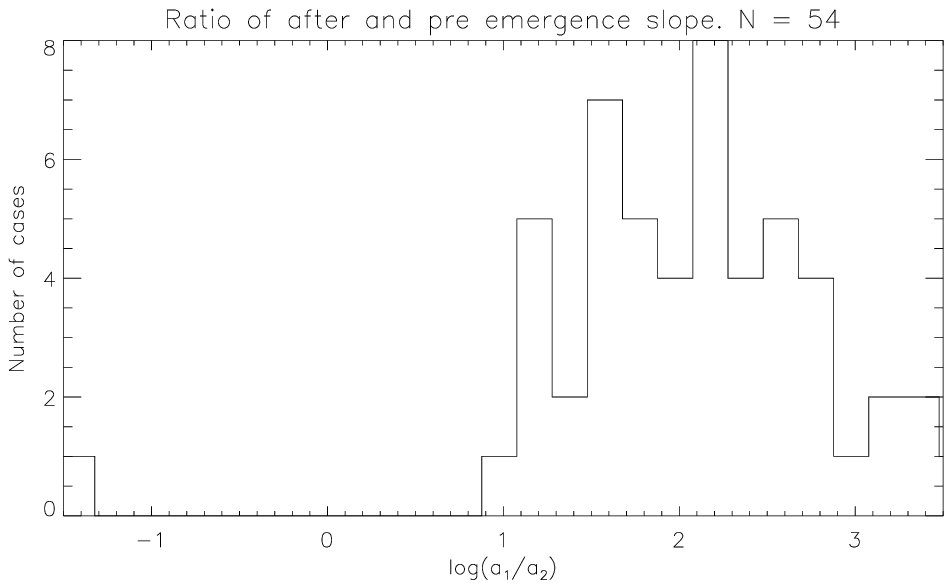, width=4.0in, clip=true}}
\caption{\footnotesize Top: The histogram of time differences between AR
  appearance times from the SRS reports and emergence times that we
  estimate from our fitted break point for the gradual-to-rapid
  transition in the unsigned flux curve.  Positive times correspond to
  delayed appearance in the SRS reports.  This histogram includes 104
  active regions.  Middle: 
  The histogram of the fraction of total unsigned flux at the inferred time of
  emergence to the maximum total unsigned flux of each active region: red
  histogram is for our fitted emergence times, and purple is for
  appearance times in SRS reports. The dotted line shows the 10\%
  threshold adopted by \citet{Leka2013}.  We see that in many cases,
  the fitting procedure determines emergence times prior to the 10\%
  threshold.  In contrast, SRS appearances are almost equally likely
  to occur at any fraction of a new region's maximum flux. Bottom: This
  histogram shows the ratio between the rapid-phase and gradual-phase
  rates of flux emergence, on a natural log-linear axis, so greater
  than 0 means larger slope after emergence. This distribution peaks
  around 2.2, implying $a_2/a_1 \gsim 7$. }
\label{fig:fluxanal}
\end{figure}

We deemed our fitted slope break points to accurately capture the
emergence time in 104 cases. (In one of the 105 well-fitted cases, the
fitted lines approximately matched the overall flux-versus-time
evolution, but the break point in slopes found by our algorithm
occurred at a time that differed from the onset of clear emergence.)
In the top panel of Figure \ref{fig:fluxanal}, we plot a histogram of
the time lags between our emergence time and the first appearance in
the SRS reports, with positive corresponding to an earlier detection
by our algorithm.
A Gaussian fit (overplotted in blue) is centered at 26 hr, with a sigma of 
15 hr.
The distribution peaks near +24 hr, meaning our algorithm tends to
define emergence times about a day ahead of regions' appearances in
SRS reports.
This confirms a similar result reported by \citet{Leka2013}.
The distribution's standard deviation is 12 hours, which shows that
the SRS time is good as long as we do not go beyond its time
resolution of one day. 
In Tables \ref{tab:slopes1} and \ref{tab:slopes2}, we list the SRS and
fitted appearance times for the 104 well-fitted regions, in which our
fitting algorithm's break point in slopes matched the onset of
emergence.  We also list the initial and final slopes during the
fitting interval.

\citet{Leka2013} suggested a different definition for the emergence
start time: the point at which a region's total flux is 10\% of the
maximum over the lifetime of the region. However, at least with our
imperfect computation of total flux, this 10\% rule for many regions
does not select a meaningful point for the start of emergence.
As Figure \ref{fig:fluxplot} shows, 10\% of the
maximum flux may appear very early, when the uncertainty in the total
flux is relatively large.
Further, the exact maximum of a region's unsigned flux can also be
somewhat uncertain.

In the middle panel of Figure \ref{fig:fluxanal}, we plot the
histogram of the fraction of each region's flux present at two
possible definitions of appearance time (cf., emergence time): our
break-point approach (in red) and the first-SRS time (in purple).
Because a new region can appear at any time, but SRS reports are
  only issued once per day, the fraction of peak flux present at the
  first-SRS time is essentially random.  In contrast, at the majority
of two-slope break-point appearance times, less than 10\% of regions'
flux has emerged.  For comparison, the 10\% cutoff is shown as a
vertical, dashed line.
%

In the bottom panel of Figure \ref{fig:fluxanal}, for two-phase
events, we plot a histogram of the natural logarithms of ratios of
slopes in the first and second phases, $a_1$ and $a_2$, respectively.
The distribution is broad, with peak at $\log(a_2/a_1) = 2.2,$
corresponding to a ratio of slopes of more than 7 to 1 between the
rapid and gradual phases.

While we believe the emergence times acquired from curve fitting are
more accurate than those from SRS reports, NOAA's SRS database
provides a much larger amount of cases for statistical analyses.  This
is especially true when applying various thresholds on new active
regions to consider, e.g., region size, or proximity to the nearest
PEAR.  We chose not to combine our set of fitted times with the
larger set of SRS times, because doing so would make our emergence
times inconsistent.

\subsection{Superposed Epoch Analysis of Flares and Emerging ARs}
\label{subsec:epoch}


To quantitatively characterize the relationship between active region
emergence and flaring in PEARs, we perform superposed epoch analysis
on active region emergence events and flares. The idea of superposed
epoch analysis (or Chree analysis) is to co-register the times of one
set of events at a common, ``key'' time, and superpose the event
counts for the other set of events relative to the co-registered key
time. A peak at or near the key time would indicate a relation between
these two sets of events. We define the key time to be the emergence
of new active regions, and compute rates of flaring in PEARs relative
to this key time.

For emergence times, we use the SRS reports: we treat a region's first
appearance in an SRS report as its emergence time.  For regions that
first appeared within 45 degrees of disk center, we then investigate
flare activity in PEARs $\pm 72$ hours from this time using
the GOES flare 
catalog.\footnote{ftp://ftp.ngdc.noaa.gov/STP/space-weather/solar-data/solar-features/solar-flares/x-rays/goes/}
The maximum umbral area of new regions (also given in the SRS reports,
with correction for foreshortening) is expected to be correlated with these
regions' total unsigned fluxes (e.g., Figure 4 of
\citealt{Fisher1998}), and should thus be related to the strength of
interaction between the new region and PEARs. New AR size therefore
provides a plausible criterion to select sets of active regions to
investigate for a dose-response relationship.

\begin{figure}[htb]
\psfig{figure=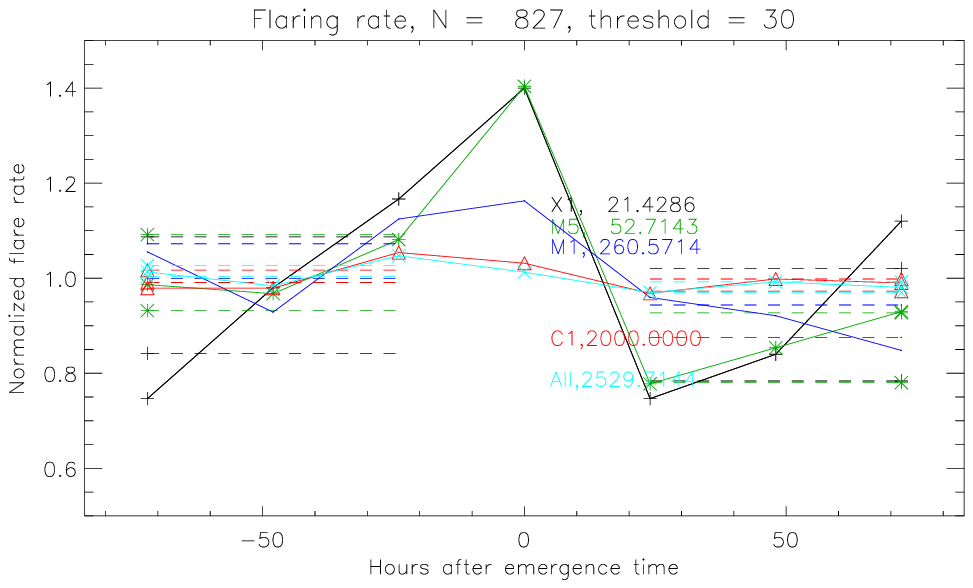, width=4.5in, clip=true }\\
\psfig{figure=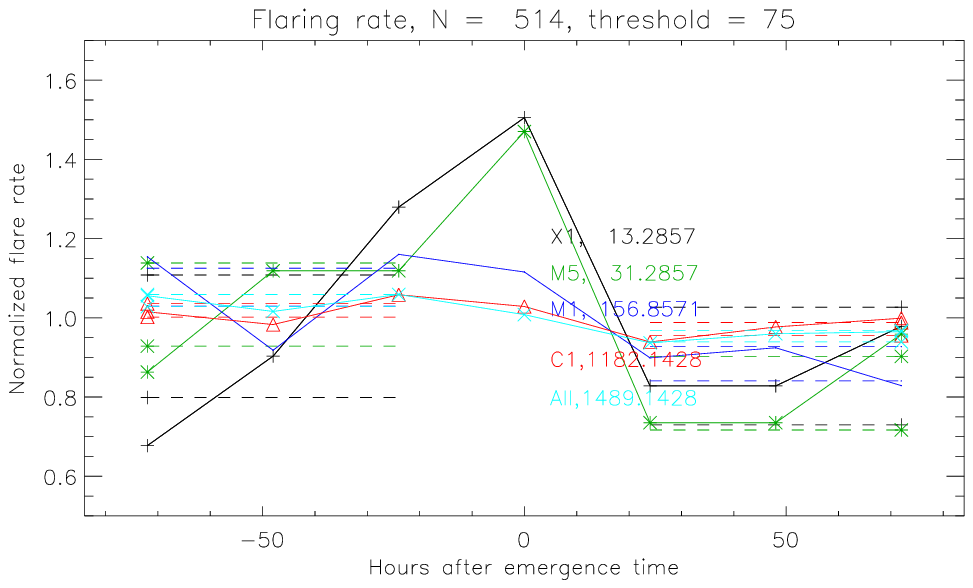, width=4.5in, clip=true}\\
\psfig{figure=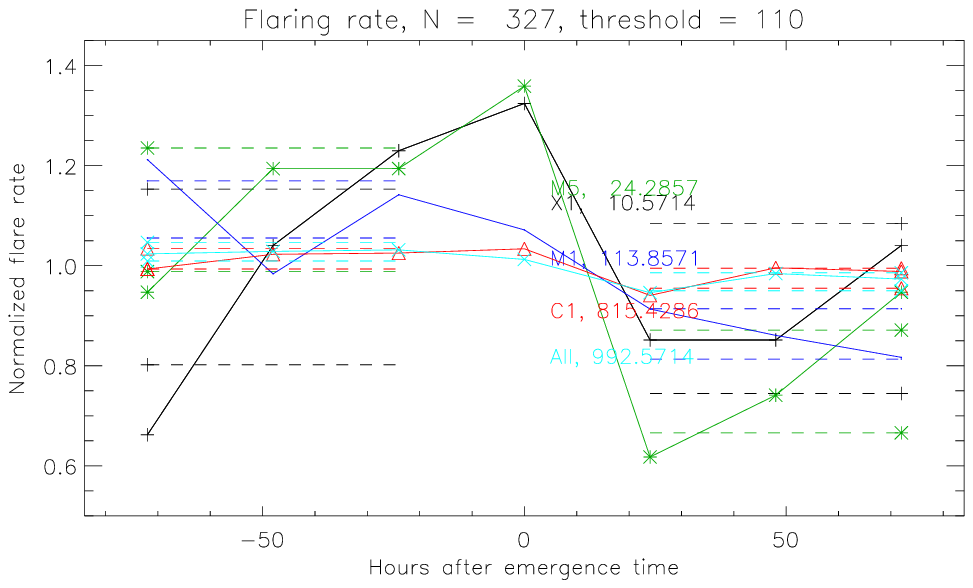, width=4.5in, clip=true}%
\caption{\footnotesize These plots show rates of flares in PEARs
  before and after the key time determined by new active region
  emergence. On each plot, each data point shows the normalized flare
  count rate, for several flare thresholds, summed over all key times,
  with each flaring rate scaled so that the mean rate is 1 over the 7-day
  interval shown.
  Symbols differ for each flare threshold: they are $\times,
  \triangle,$ dot (.), $*,$ and $+$, for all flares, C1,
  M1, M5, and X1, respectively.
  The actual mean rate (in units of flares per day per N emergence
  events) is annotated for each line.
  The dashed lines, marked by color and endpoint symbols, are mean 
  rates $\pm$ 1-$\sigma$
  uncertainty intervals for the 3 days before or after the key time,
  excluding the key-time interval itself
  (to better represent natural variations in the rate 
  without the influence of the new active region). 
  From top to bottom, only regions larger than increasing thresholds
  for new active region size (in micro solar hemispheres, or MSH) are
  included.}
\label{fig:flrrat1}
\end{figure}

\begin{figure}[htb]
\psfig{figure=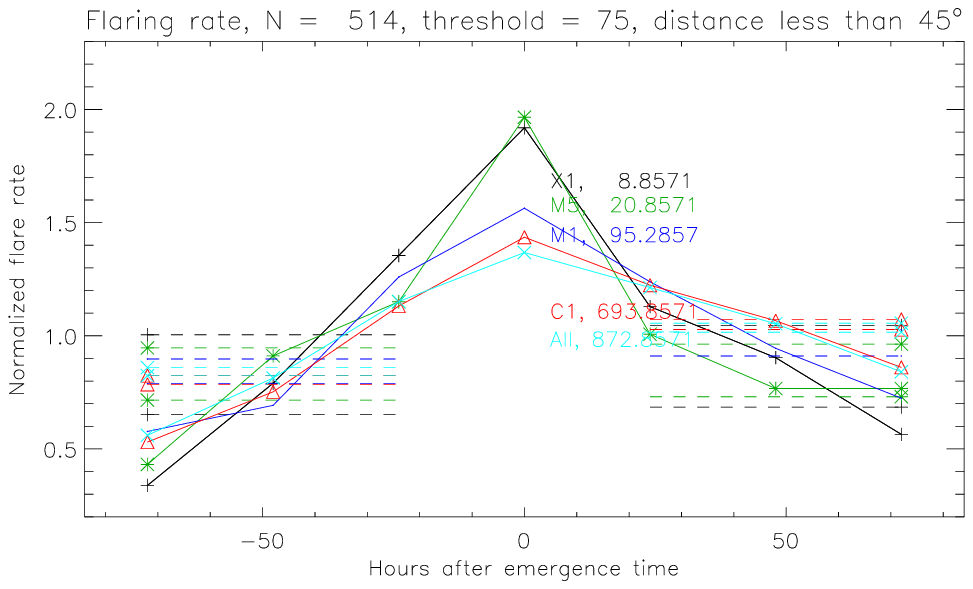, width=4.5in, clip=true}\\
\psfig{figure=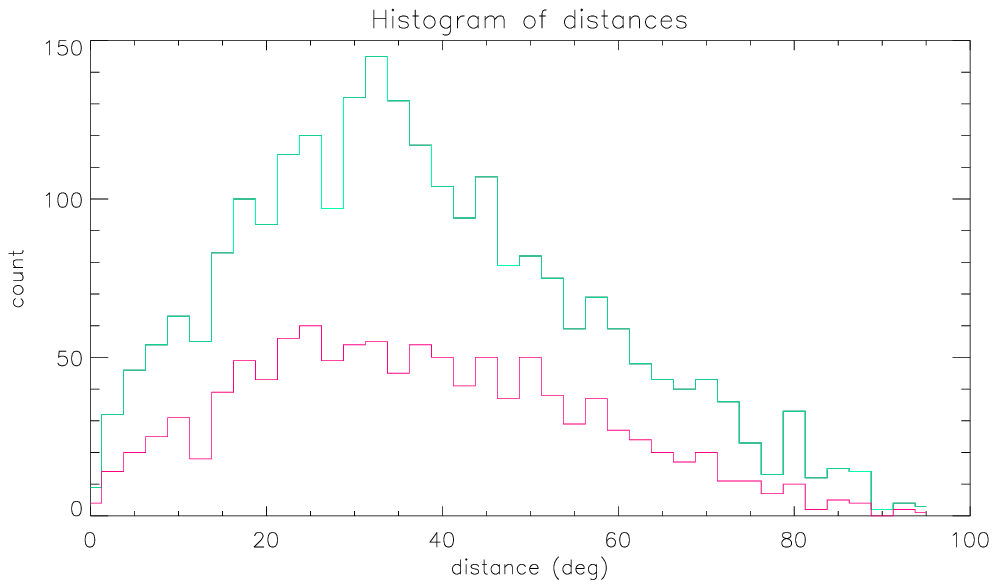, width=4.5in, clip=true}\\
\psfig{figure=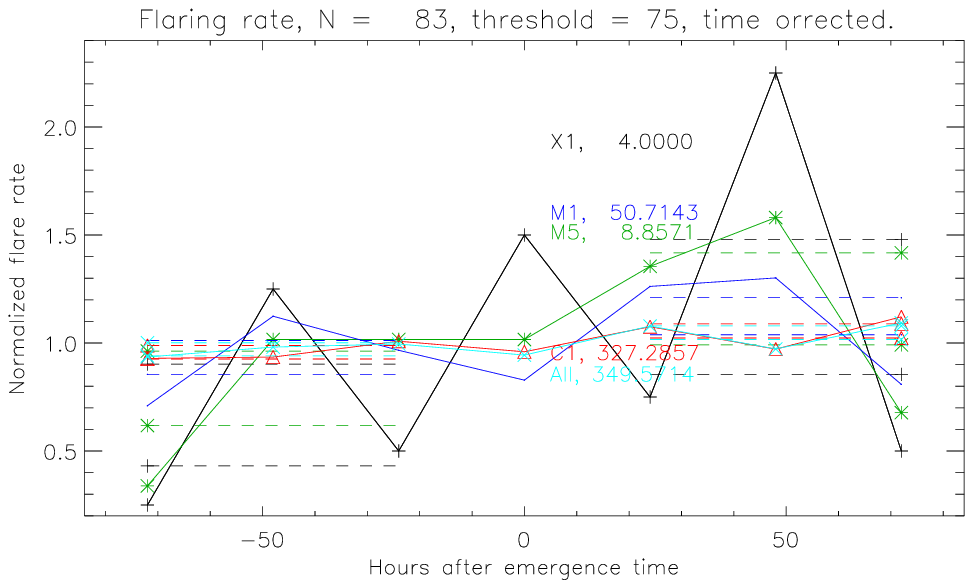, width=4.5in, clip=true}%
\caption{\footnotesize Top: rates of flares in PEARs, as in the
  previous figure, but including only flares from PEARs closer than
  45 degrees from the newly emerged region's location.
  Middle: histograms of flare distances (green) from each new
  region, and distances (red) to the nearest PEAR for each emergence.
  Bottom: flare rates of PEARs in the subset of active regions for
  which we determined an emergence time from the break-point
  approach. }
\label{fig:flrrat2}
\end{figure}

Figures \ref{fig:flrrat1} and \ref{fig:flrrat2} give our primary
results regarding the influence of new regions on flaring in PEARs.
In each panel of Figure \ref{fig:flrrat1}, flaring rates for several
GOES flare thresholds
(all flares, C1, M1, M5, and X1) 
are shown with solid lines for seven 24-hour periods: $\pm 12$ hours
from the first SRS report listing a new region, and $\pm$ 3 days on
either side of this.
These rates are in units of flares per day per $N$ emergence events,
where $N$ is given at the top of each plot: the rate is computed by
simply summing all flares for all emergence events, and dividing by
seven.
To fit rates for all thresholds on a single plot, the mean rate for
each flare threshold over the entire 7-day interval was then scaled to
one; the unscaled daily-average flare rate is given for each class
next to that class's label.
Dotted lines show the mean rate $\pm$ 1-sigma expected uncertainty
ranges (see below) in the flaring rate for each threshold, assuming
that Poisson statistics govern the count rate, computed separately for
the 3-day time periods before and after the key-time interval (and
excluding the key-time interval itself).

To search for any evidence that PEAR flare rates depend upon
new-region size, we computed their flare rates for three subsets of
emergence events, with increasingly restrictive minimum thresholds on
new regions' umbral-area sizes.  The top, middle, and bottom panels in
Figure \ref{fig:flrrat1} show PEAR flare rates associated with
emergence of new regions with peak umbral areas larger than 30, 75,
and 110 micro solar hemispheres (MSH) or microhemispheres (mhs),
respectively.
While we expected larger new regions to more strongly affect PEARs
flare rates than smaller new regions, we do not see clear evidence of
this effect.
One possible explanation for this lack of strong size dependence is
that flare-relevant perturbations to PEARs' magnetic environments
might set in after only modest amounts of flux have emerged, so the
later growth of regions to their peak sizes does not matter.

In all three panels of Figure \ref{fig:flrrat1}, the fractional
increase in the rate of more powerful flares is larger than increases
in the rates of weaker flares.
One possible explanation is obscuration, an effect pointed out by
\citet{Wheatland2001}, who showed that the GOES catalog suffers from
the tendency of the flare identification procedure to fail to identify
relatively small flares that follow large flares.  Small flares are
also lost in the GOES background during very active periods
(e.g., \citealt{Hudson2014}).
%
Another plausible explanation is that the higher peak GOES X-ray flux
necessary for a flare to be characterized as more powerful likely
requires the involvement of magnetic fields over a larger area in the
flare.  Physically, this might mean more magnetic flux reconnects;
observationally, this would in many cases produce flare ribbons
sweeping over more photospheric magnetic flux, and post-flare loops
over a larger area.  Hence, larger flares are likely more non-local,
i.e., they probably involve more global-scale magnetic connections
relative to smaller flares.
It can also be seen that the increases in flaring rates precede the
time of the first SRS report by a few dozen hours, consistent with the
delay in SRS reports noted above (and shown in the histogram in the
top panel of Figure \ref{fig:fluxanal}).

We included flares from PEARs any distance from new regions when
computing the rates shown in Figure \ref{fig:flrrat1}.  
To investigate whether PEARs in close proximity to new regions are
more likely to flare, in the top panel of Figure \ref{fig:flrrat2}, we
plot PEAR flare rates for the subset of new regions that both (i) had
peak umbral areas above 75 MSH in size and (ii) were less than
45$^\circ$ from nearest PEAR.
The peaks at the key time in this panel are larger than the peaks in
the middle panel of Figure \ref{fig:flrrat1} (which has the same
region-size threshold), implying that closer emergence events are more
strongly assocated with increased flare activity.
This is in accordance with expectations from our interaction-energy
model, based upon the scaling of the potential field interaction with
distance.
Note also that for this set of PEARs close to the emerging region,
the rates of {\em all} classes of PEAR flares increase.
We remark that NOAA's AR designations are somewhat subjective; some
observers might characterize a new NOAA region close to a PEAR as
emergence associated with that PEAR instead.
To investigate this idea, in the middle panel of Figure
\ref{fig:flrrat2}, we plot histograms of distances to (i) PEAR flares
of any class (green) and (ii) the nearest PEAR to the emergence of any
region larger than 75 MSH.
Both PEARs and flares are quite close to new regions in some cases,
but in the majority of cases they are clearly separated.

%
%
%
In the bottom panel of Figure \ref{fig:flrrat2}, we plot PEARs' flare
rates for the subset of new regions for which: (i) the break point in
our two-phase linear fits to the flux-versus-time curves were deemed
to match the region's emergence time; and (ii) areas exceeded 75 MSH.
Compared to the humps in flare rates found from the SRS appearance
times, the onset of the flare rate increase shifted to the right,
consistent with our expectation that the SRS appearance times are later
than the true emergence by about 24 hr.
Also, the peak for flares X1 and greater is higher than for the
corresponding SRS-emergence plot (middle panel of Figure
\ref{fig:flrrat1}), which is consistent with more accurate estimation
of the key times in our superposed epoch analysis.

\begin{figure}[htb]
\psfig{figure=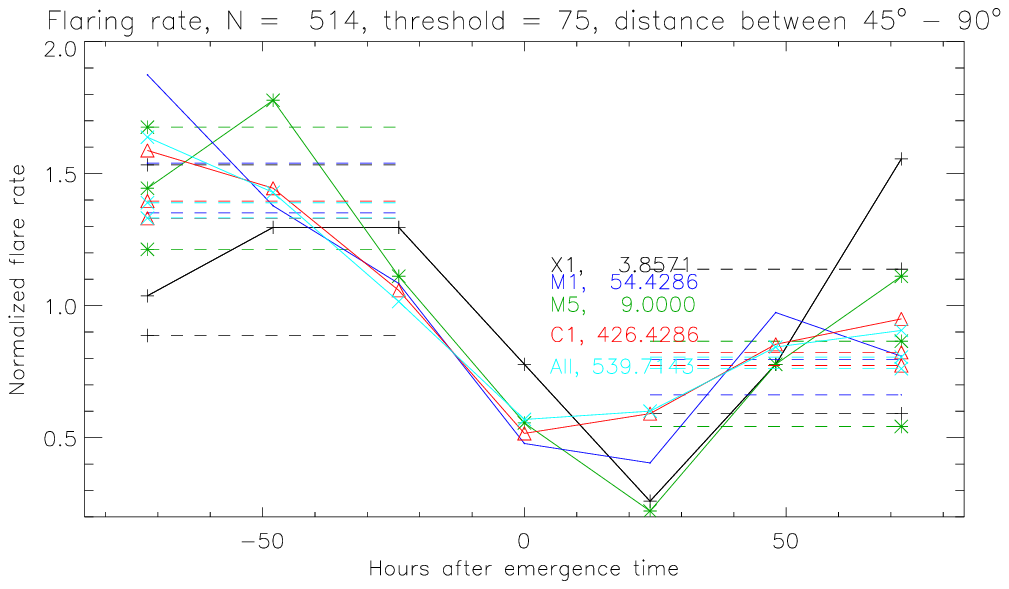, width=3.75in, clip=true} \\ %
\psfig{figure=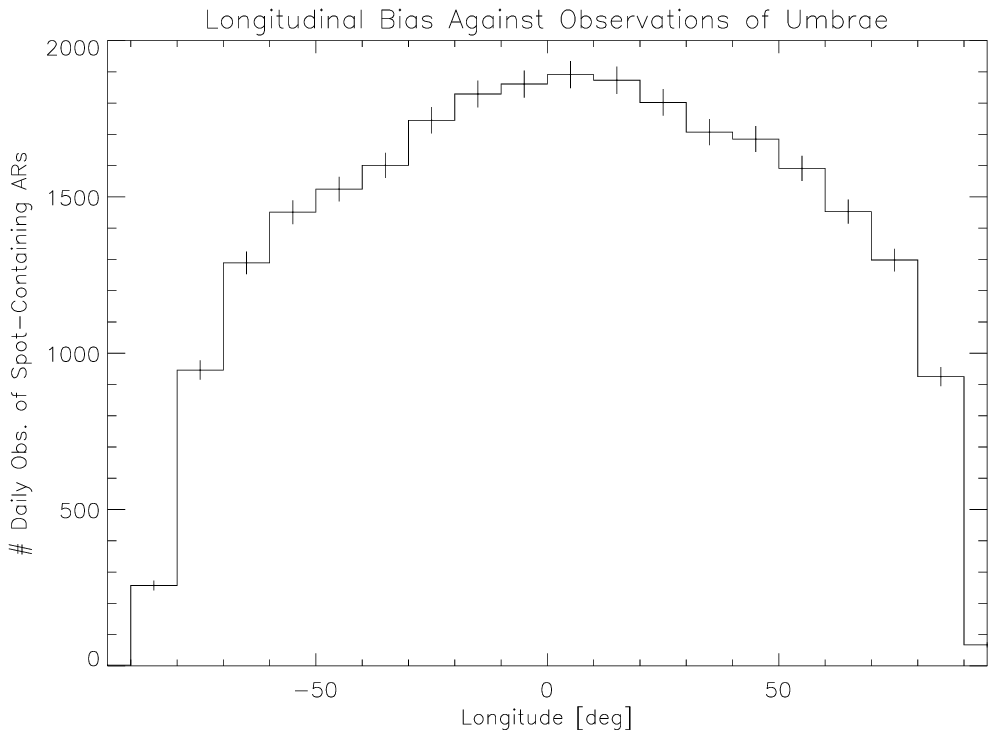, width=3.75in} \\ %
\psfig{figure=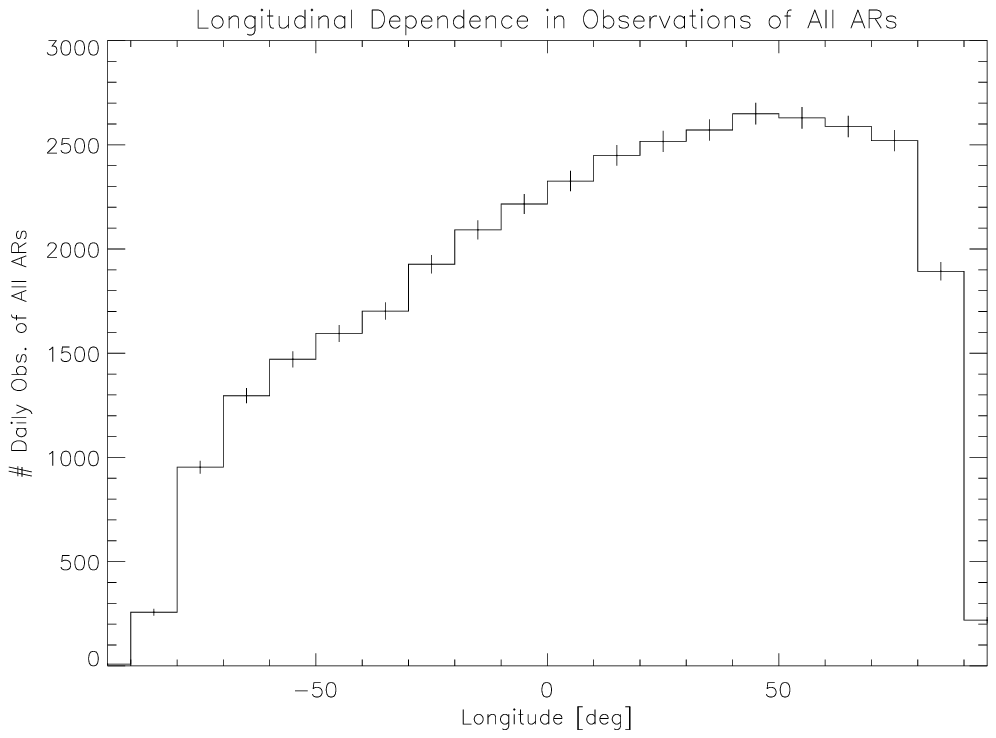, width=3.75in}
\caption{Rates of flares in PEARs between 45$^\circ$ and 90$^\circ$
  from new regions.
  It exhibits a ``U'' shape, which can be understood from the
  distribution of recorded locations of spot-containing regions
  (middle panel), which shows fewer regions toward the limbs; for a
  point near disk center, there are relatively few PEARs $> 45^\circ$
  away, so the PEAR flare rate is lower.
  There is also an asymmetry in the top plot, with the left side
  higher, corresponding to more flares originating in west-disk PEARs.
  This is consistent with E-W biases (e.g., \citealt{Dalla2008}) in the
  distributions of both spot-containing regions (middle panel) and all
  NOAA regions (bottom panel; this includes H$\alpha$ plages without
  spots): fewer regions are identified on the eastern disk, so fewer
  east-disk flares are likely to be ascribed to any region.
}
\label{fig:far-bias}
\end{figure}

While the increase in flares in PEARs closer than 45$^\circ$ is
evident, the rate on PEARs farther away is less clear.  In the top
panel of Figure \ref{fig:far-bias}, we show the rates of flares in PEARs
between 45$^\circ$ and 90$^\circ$ from new emergences.
To keep a uniform area that excludes beyond-limb regions, we only
include regions in the half of the 90-degree circle centered on the
new region that lies toward disk center and is bisected by the new
region's meridian.
This implies that the initial three days of most regions' histories
involve westward PEARs, while the final three days of most regions'
histories involve eastward PEARs.
The ``U'' shape of the rate curves can be understood from the
distributions of recorded locations of spot-containing regions (middle
panel), which show a relative dearth of regions toward the limbs: for
points toward $75^\circ$ in either E or W longitude, there are more
regions 45$^\circ$ - 90$^\circ$ away than there are for points near
disk center.  Hence, over each 7-day interval used in our superposed
epoch analysis, the likelihood of flares from PEARs 45$^\circ$ -
90$^\circ$ away is highest at points away from disk center.
There is also an asymmetry in the top plot, with the left side higher,
corresponding to more flares originating in west-disk PEARs.  This is
consistent with E-W biases in the distributions of both
spot-containing regions (middle panel) and all NOAA regions (bottom
panel, which also includes H$\alpha$ plage regions lacking sunspots):
fewer regions are identified on the eastern disk, so fewer east-disk
flares are likely to be ascribed to any region.
This E-W bias was previously reported by \citet{Dalla2008}.

\subsection{Significance of Flaring Rate Variations}
\label{subsec:erranal}

To characterize the significance of variations in flaring rates
associated with the emergence of new regions, we estimate the random
rate variations that would be expected under the null hypothesis,
i.e., that the emergence of new regions does not have any effect on
flaring in PEARs. To do so, we separately estimated the flare rate and
its uncertainty for the three-day intervals before and after the key
time.  Within each three-day interval, we treat the $j$-th day's flare
count, $n_j$, as a separate estimate of the average flare rate
$\langle n_f \rangle$ per day per $N$ emergence events, with
\begin{equation}
  \langle n_f \rangle = \frac{1}{J} \sum_{j=1}^{J} n_j 
~,
\label{eqn:rate} \end{equation}
for observations over $J$ days.
Given the delay in appearance times in NOAA SRS reports relative to
emergence times that we found from flux-versus-time curves (see the
top panel of Figure \ref{fig:fluxanal} above), we expect that our
pre-appearance flaring rate for SRS key times will be partially
``contaminated'' by new-region flux.  To the extent that emergences do
indeed increase the PEAR flaring rate, this contamination will
increase the apparent background PEAR flaring rate, complicating the
task of showing an increase above background associated with the new
region.  This is a price we have agreed to pay to increase our sample
size with the SRS dataset.

We then assume Poisson statistics govern fluctuations in each day's
estimate of the flare rate, i.e., given $n_j$ flares on day $j$, the
standard deviation in that day's rate estimate is $\sqrt{n_j}$.  For
$J$ successive days of observations, we take the standard error in the
mean flare rate, $\sigma_f$,
\begin{equation}
\frac{1}{\sigma_f^2} = \frac{J}{\langle n_f \rangle} ~,
\label{eqn:sigma} \end{equation}
%
%
as the uncertainty.

These averages and standard deviations in PEAR flare rates are used in
Figures \ref{fig:flrrat1} and \ref{fig:flrrat2}. In all plots of flare
rates, solid lines show the flare rate 
for each flare threshold, normalized to make each threshold's mean
daily rate over the full, 7-day interval equal to unity.
The dashed lines show $\langle n_f \rangle \pm \sigma_f$, where
$\langle n_f \rangle$ and $\sigma_f$, where computed separately over
the three-day pre- and post-emergence intervals.

We computed separate pre- and post-emergence uncertainty estimates on
the flare rates for two reasons.
First, as noted above, the GOES catalog suffers from obscuration.
If new-region emergence does trigger flares in PEARs, some of which
are large, then the subsequent PEAR rate for smaller flares will be
systematically diminished by obscuration.
Second, on physical grounds, the flare rate in PEARs might be
suppressed after a large flare happens, since the large-scale,
post-flare magnetic field should be in a more-relaxed state.  We refer
to this possible phenomenon as ``flare shadows,'' based upon the
analogous effect in terrestrial geology, ``stress shadows'' (e.g.,
\citealt{Mallman2008}), in which most faults near one that just
produced an earthquake are de-stressed.  In the coronal magnetic
field, separators play the role of terrestrial faults.
While \citet{Wheatland2001} showed convincing evidence that
obscuration affects flare counts in the GOES catalog, this by itself
does not rule out the existence of flare shadows.

%
%

Although essentially all of our flare rate plots exhibit noticeable
peaks near the key time for M- and X-class flares, most peaks are
basically 2$\sigma_f$ or less above the mean rate.  Exceptions are for
X1 and M5 or greater flares from any PEARs, {\em all} classes of
flares from close PEARs (top panel of Figure \ref{fig:flrrat2}), and
for X1 or greater flares in PEARs for the subset of new regions in
which we estimated the emergence time from the flux-versus-time curve
(bottom panel of Figure \ref{fig:flrrat2}), all of which are near
3$\sigma_f$ enhancements.
We therefore characterize the effect for large flares as statistically
significant: the emergence of new regions is associated with
increases in the rate of large flares in all PEARs, and the rate
of all classes of flares for PEARs close to the emerging regions.
Given the relatively small numbers of large flares, however, the
possibility that the peaks that we see arose from statistical
fluctuations must be kept in mind.

\section{Conclusions}
\label{sec:conclu}

In this paper we studied the influence of newly emerging active
regions on flaring rates in PEARs.  

We first presented a theoretical approach, based upon an interaction
energy between new and pre-existing regions defined in terms of pre-
and post-emergence potential fields.  This interaction energy depends
upon the part of potential field energy arising from coupling between
pre-existing flux systems and the new active region. This term should
generally not be a trivial replication of a new region's total
unsigned flux, as we have shown in case studies of two regions, AR
10488 and AR 11158, that emerged near pre-existing regions.
We argue that, based upon Hale's law, this interaction energy should
often be negative, and that when negative it implies the presence of
free energy in the coronal field as a result of the new region's
emergence.
We use the term topological free energy to refer to free energy
associated with a negative interaction energy between a new region and
PEARs, as distinct from the internal free energy within each flux
system.
In our two case studies, the system we found to possess topological
free energy, the AR 10488/AR 10486 pair, the nearby PEAR (10486)
produced a large flare; our approach did not find topological free
energy in the other system (AR 11158/11156), and the nearby PEAR
(11156) did not produce any large flare.

We found, however, that even when a big, new region emerges very near
another large PEAR, the interaction energy turns out to be relatively
small compared to the energy released in large flares. Thus we
conclude that this interaction energy cannot directly explain the
energy released in subsequent flares.  Instead, we suggest that the
interaction energy might quantify the strength of the perturbation to
the magnetic environment of PEARs --- and if a given PEAR is only
marginally stable, a larger interaction energy should imply a greater
likelihood of a flare.  
In this view, new region emergence might trigger flares or CMEs by
processes such as breakout reconnection \citep{Antiochos1999a}
between overlying fields in the old flux system and the new flux
system, enabling stressed, inner fields in the old flux system to
erupt.
We believe that more involved analyses of the magnetic interactions
between new and old regions using potential fields --- such as those
undertaken by \citet{Longcope2005a}, \citet{Tarr2012}, and
\citet{Tarr2014} --- can provide much more information than our
interaction-energy approach, although only with greater effort.
%

We then undertook a statistical study of the rates of flares in PEARs
near the times at which new regions emerged, using superposed epoch
analysis.
Our results suggest that the emergence of new regions is
  associated with increases in the rate of flares in PEARs,
especially M- and X-class flares, with the size of the effect on the
flare rate being 2$\sigma_f$ or more.
Relatively little influence of new-region emergence on the rates of
smaller flares in PEARs was found, except for PEARs within 45$^\circ$ of
new regions.  This is perhaps because smaller flares might arise
from local magnetic field structure and processes that are less
influenced by changes in the large-scale structure of the coronal
magnetic field.
\citet{Dalla2007} investigated flare rates associated with new regions
that emerged within 12$^\circ$ of a PEAR, but found only a relatively
small increase in the flare rate averaged over the 4 days following
the emergence times.  We note that the rate increases that we find
using SRS reports typically start about one day before this
``emergence time,'' and only last about one day afterward.  Hence, the
effects we see could be substantially diminished in 4-day averages of
post-SRS flare rates.


We believe that the statistical association between new-region
emergence and temporary enhancements in the rate of remote flaring
that we report arises from large-scale magnetic interactions between
each new region and PEARs in the corona.  Our observations, however,
only demonstrate correlation, not causation.  So it is possible that
some other physical process(es) explains the correlation.

One possible alternative explanation is the organization of magnetic
fields at super-active-region scales, in ``activity complexes''
\citep{Bumba1965} or ``activity nests'' \citep{Schrijver2000}.
It is well known that new flux is much more likely to emerge within
PEARs (e.g., \citealt{Liggett1985, Harvey1993}), and observations show
that flares within a PEAR tend to be associated with emergence
of additional flux within that PEAR (e.g., \citealt{Martin1982}).
It is possible that there exists some physical connection, through the
solar interior, between the emergence of (i) additional within PEARs
and (ii) new flux that forms a new AR.
Such coordinated emergence could produce correlations like those we
report.
The necessary length scale of such super-active-region connections can
be inferred from the red histogram in middle panel of Figure 8, which
shows the distribution of distances to the nearest PEAR for each
emergence.  The bulk of new emergences in our are farther than 15
heliocentric degrees (i.e., greater than 180 Mm) from pre-existing
regions.

In our view, observations of enhanced flare activity in PEARs due to
the emergence of new regions have clear implications for understanding
the release of magnetic energy in flares and CMEs.  Foremost,
interactions between coronal magnetic flux systems are often global in
scale: evolution in one part of the corona can drive evolution in
distant regions.  This idea is not new, and has been discussed by many
other researchers.  Examples include the contexts of
emergence-triggered filament disruptions (e.g., \citealt{Bruzek1952,
  Feynman1995}), transequatorial loop systems observed with the soft
X-ray telescope aboard the Yohkoh satellite \citep{Pevtsov2000}, PFSS
models of the implications of AR orientation with respect to the
background coronal field \citep{Luhmann2003}, and sympathetic flaring
(e.g., \citealt{Moon2002d, Schrijver2011b}).  Our results are another
manifestation of this global-scale coupling. 

These considerations are clearly relevant for space weather
forecasting, since far-side evolution might have implications for
front-side events, which can be geoeffective: {\bf What you don't know
  \underline{can} hurt you!}  In particular, geoeffective front-side
flares and CMEs might be arise from processes associated with
beyond-limb flux emergence.
Currently, new-region emergence on the far side can be detected and
calibrated in terms of total unsigned flux using helioseismology
(e.g., \citealt{Hernandez2007}) and / or by combining STEREO EUV
observations of emergence (e.g., \citealt{Liewer2014}) with known
scalings between magnetic flux and EUV radiance \citep{Fludra2008}.
Our results motivate satellite missions enabling full-Sun (or
``4$\pi$'') observations of the solar surface and atmosphere, for
reasons both practical (to better predict space weather) and
scientific (to better understand large-scale magnetic connections and
their implications).
In addition, for simulation efforts to understand active-region
evolution (e.g., \citealt{Cheung2012}), neglect of active regions'
global magnetic environment might ignore the crucial role of
global-scale connections.

Beyond our main results, our efforts to objectively estimate active
region emergence times from magnetograms led to an additional finding:
many active regions (about half in our sample) exhibit ``two-phase''
emergence, with a phase of relatively gradual emergence followed by a
relatively sudden change to more rapid emergence.
We find that a piece-wise linear fit to the time profile of total
unsigned magnetic flux captures this transition in the dynamics of
emergence in many regions.
Two-phase emergence has been seen in simulations of flux emergence, as
in the modeling undertaken by \citet{Toriumi2011}, who found that the
rate of emergence in a single emerging flux structure varied with time
as the structure reached the photosphere.  In contrast,
\citet{Chintzoglou2013} believe the gradual-to-rapid emergence in
their observations of AR 11158 arose because the emerging structure
was fragmented, with a weaker-field structure preceding a
stonger-field structure.  Further study of flux-versus-time profiles
in models and observations is warranted.

While previous authors \citep{Leka2013} defined emergence times from
the appearance of 10\% of the (subsequent) maxima of regions' total
unsigned flux, we argue that defining the emergence time from the
fitted emergence rate --- and perhaps at the transition point from
gradual to rapid --- is more robust to fluctuations (from noise or
other artifacts) in the observed total unsigned flux, and captures an
important physical aspect of the emergence process.
As another by-product, we found reports of new active regions in
NOAA's Solar Region Summaries typically occurred about 24 hours after
the actual start of significant new flux emergence, in agreement with
previous reports \citep{Leka2013}.

\acknowledgements 
%
We thank the referee for providing a detailed review of our
manuscript, and for valuable suggestions to improve it, including the
suggestion of super-active-region connections within the solar
interior.
We thank the American taxpayers for supporting this work.
We are also grateful to Hugh Hudson, Mike Wheatland, and Sylvia Dalla
for reading a draft of this paper, and offering suggestions that
improved it.
We acknowledge funding from the NSF's National Space Weather Program
under award AGS-1024862, the NASA Heliophysics Theory Program (grant
NNX11AJ65G),  and the Coronal Global Evolutionary Model (CGEM) award
NSF AGS 1321474.



\end{document}